\documentclass[reprint, notitlepage,superscriptaddress,preprintnumbers,nofootinbib,nobibnotes,amsmath,amssymb,aps,twocolumn]{revtex4-1}
\usepackage{graphicx,grffile}
\usepackage{dcolumn}
\usepackage{bm}
\usepackage{hyperref}
\usepackage{aas_macros,braket}
\makeatletter
\def\lst@lettertrue{\let\lst@ifletter\iffalse}
\makeatother

\begin{document}
\title{Detecting higher spin fields through statistical anisotropy in the CMB and galaxy power spectra}
\author{Nicola Bartolo} 
\affiliation{Dipartimento di Fisica e Astronomia ``G. Galilei'', Universit\`a degli Studi di Padova, via Marzolo 8, I-35131, Padova, Italy}
\affiliation{INFN, Sezione di Padova, via Marzolo 8, I-35131, Padova, Italy}
\affiliation{INAF-Osservatorio Astronomico di Padova, vicolo dell Osservatorio 5, I-35122 Padova, Italy}
%---
\author{Alex Kehagias} 
\affiliation{Physics Division, National Technical University of Athens, 15780 Zografou Campus, Athens, Greece}
%---
\author{Michele Liguori}
\affiliation{Dipartimento di Fisica e Astronomia ``G. Galilei'', Universit\`a degli Studi di Padova, via Marzolo 8, I-35131, Padova, Italy}
\affiliation{INFN, Sezione di Padova, via Marzolo 8, I-35131, Padova, Italy}
\affiliation{INAF-Osservatorio Astronomico di Padova, vicolo dell Osservatorio 5, I-35122 Padova, Italy}
%---
\author{Antonio Riotto}
\affiliation{Department of Theoretical Physics and Center for Astroparticle Physics (CAP) 24 quai E. Ansermet, CH-1211 Geneva 4, Switzerland}
%---
\author{Maresuke Shiraishi}
\affiliation{Department of General Education, National Institute of Technology, Kagawa College, 355 Chokushi-cho, Takamatsu, Kagawa 761-8058, Japan}
%---
\author{Vittorio Tansella}
\affiliation{Department of Theoretical Physics and Center for Astroparticle Physics (CAP) 24 quai E. Ansermet, CH-1211 Geneva 4, Switzerland}
%---

\date{\today}

%\pacs{98.80.Cq}

\begin{abstract}
%\noindent

  Primordial inflation may represent the most powerful collider to test high-energy physics models. In this paper we study the impact on the inflationary power spectrum of the comoving curvature perturbation in the specific model where  massive higher spin fields are rendered effectively massless during a de Sitter epoch through suitable couplings to the inflaton field. In particular, we show that such  fields with spin $s$  induce a distinctive statistical anisotropic signal on the power spectrum, in such a way that not  only the usual $g_{2M}$-statistical anisotropy coefficients, but also higher-order ones (i.e., $g_{4M}$, $g_{6M}$, $\cdots$, $g_{(2s-2)M}$ and $g_{(2s) M}$) are nonvanishing. We examine their imprints in the cosmic microwave background and galaxy power spectra. Our Fisher matrix forecasts indicate that the detectability of $g_{LM}$ depends very weakly on $L$: all coefficients could be detected in near future if their magnitudes are bigger than about $10^{-3}$. 

\end{abstract}

\maketitle

%~~~~~~~~~~~~~~~~~~~~~~~~~~~~~~~~~~~~~~~~~~~~~~~~~~~~~~~~~~~~~~~~~~~~~~~~~~~~

%%%%%%%%%%%%%%%%%%%%%%%%%%%%%%%%%%%%%%%%%%%%%%%%%%%%%%%%%%%%%%%%%%%%%%%%%%%%%%
\section{Introduction}
%%%%%%%%%%%%%%%%%%%%%%%%%%%%%%%%%%%%%%%%%%%%%%%%%%%%%%%%%%%%%%%%%%%%%%%%%%%%%%

%\noindent

Inflationary models for the early universe fit very well with various cosmological observations, and in particular those related to the cosmic microwave background (CMB) anisotropies~\cite{Ade:2015lrj,Ade:2015ava,Ade:2015xua}. Despite its observational success, we do know very little about the finest details of the inflationary dynamics and of the physics underlying inflation. The simplest models are based on a single slow-rolling scalar field, and built on an isotropic and homogeneous Friedmann-Robertson-Walker background spacetime. However, the very same fact that makes inflation so appealing, namely, the fact that it can be a privileged laboratory of high-energy physics, as high as $10^{14} \, \text{GeV}$, never achievable by terrestrial laboratories, urges us to be open minded on the very nature of inflation itself. For example, on the one hand, an interesting question is the following: what is the precise field content during inflation, namely, what are the extra fields other than the inflaton that could leave specific signatures in the primordial curvature perturbations generated during inflation? On the other hand, relic signatures of broken isotropy and/or homogeneity can represent a very interesting probe of inflation as well (at least to test some of the pillars on which the standard cosmological model is based on), and at the same time they can be themselves a manifestation of extra degrees of freedom (d.o.f) during inflation (signaling, e.g., the presence of vector spin-1 fields). Primordial non-Gaussianity of the perturbations arising from inflation, which nowadays is a precision test of inflation (see, e.g., the reviews~\cite{Bartolo:2004if,Chen:2010xka} and Ref.~\cite{Ade:2015ava}), can provide very useful information for both issues, possibly revealing some fine details which, e.g., can unveil the particle spectrum content of inflation via the effects of the inflaton interactions with these extra particles. 

The case of extra light scalar fields is a well-known case (see, e.g., the review~\cite{Byrnes:2010em}) as well as the case of scalar fields present during inflation with masses $m \geq H$, with $H$ denoting the Hubble parameter during inflation, which is still not so large as to simply be integrated out in the inflationary action, e.g., the so-called quasisingle field models of inflation~\cite{Chen:2009we,Chen:2009zp,Dimastrogiovanni:2015pla,Baumann:2011nk,Noumi:2012vr} (see also Ref.~\cite{Kehagias:2015jha}). References~\cite{Barnaby:2012tk,Bartolo:2012sd,Shiraishi:2013vja} showed for the first time that in the case of extra vector spin-1 fields the bispectrum of curvature perturbations $\zeta$ is characterized by a nontrivial angular structure, particularly relevant in the squeezed limit, and it was argued that this would be a specific signature of higher spin fields during inflation. In such a case, the specific angular dependence of the bispectrum is described in terms of Legendre polynomials ${\cal L}_\ell(x)$ between the three wave vectors, namely, 
\begin{equation}
B_\zeta^{n}(k_1, k_2, k_3) = {\cal C}_n {\cal L}_n(\hat{k}_1 \cdot \hat{k}_2) P_\zeta(k_1) P_\zeta(k_2) + 2~{\rm perm} , \label{eq:zeta3_cL}
\end{equation}
with $P_\zeta$ the power spectrum of curvature perturbations. Notice that in the model of Ref.~\cite{Bartolo:2012sd}, where a $U(1)$ gauge vector field is coupled to the inflaton $\phi$ field via the interaction $I(\phi) F^2$, a statistical anisotropic power spectrum and bispectrum are generated and, {\it after} an angle average, the bispectrum takes the above expression. For other models involving vector fields generating in a similar manner a bispectrum like Eq.~\eqref{eq:zeta3_cL} see, e.g., Ref.~\cite{Bartolo:2015dga}. Such an angular dependence in the bispectrum can also be generated by models of solid inflation~\cite{Endlich:2012pz} and primordial magnetic fields~\cite{Shiraishi:2012rm, Shiraishi:2012sn, Shiraishi:2013vja}. 

More recently the authors of Refs.~\cite{Arkani-Hamed:2015bza,Lee:2016vti} showed how primordial non-Gaussianity in the squeezed limit depends in a specific way on the masses and spin of higher spin particles present during inflation (see also, e.g.,~Ref.~\cite{Biagetti:2017viz}). They focused on the case of massive ($m \geq H$) higher spin particles, in which case, despite the fact that 
their fluctuations decay outside the Hubble horizon, they leave specific signatures in the correlators of the curvature perturbation $\zeta$, namely, an oscillatory behavior that depends on the masses of the extra particles and the angular structure of the primordial correlators that depend on their spin. There has been also an intense investigation of the effects on CMB and large-scale structure (LSS) of massive spin-zero particles and forecasts for their detection~\cite{Sefusatti:2012ye,Norena:2012yi,Meerburg:2016zdz,Chen:2016vvw,Ballardini:2016hpi,Chen:2016zuu,Xu:2016kwz}, along with forecasts on the angular dependence of Eq.~\eqref{eq:zeta3_cL} using various observables~\cite{Shiraishi:2013vja,Schmidt:2015xka,Munoz:2015eqa,Raccanelli:2015oma,Shiraishi:2016hjd}, and corresponding constraints from the {\it Planck} data~\cite{Ade:2015lrj,Ade:2015ava}. Only recently similar forecasts have started to be investigated for the bispectrum generated by higher spin massive particles~\cite{MoradinezhadDizgah:2017szk} to understand the detectability level of the signatures predicted in Refs.~\cite{Arkani-Hamed:2015bza,Lee:2016vti}. 

On the other hand, the results of Ref.~\cite{Bartolo:2012sd} in the case of vector spin-1 fields showed that one may expect an anisotropic background field and a large (statistical) anisotropy of the perturbations to be a general outcome of models that sustain higher than 0 spin fields during inflation. One of the crucial ingredient is a coupling of the vector field with the inflaton field to allow the spin-1 fluctuations not to decay on super-horizon scales. 
In this way a classical background vector field unavoidably gets generated at large scales during inflation from the infrared fluctuations produced during inflation. The resulting cosmological perturbations $\zeta$ that arise from such a coupling by taking into account the classical vector field  is characterized then by a breaking of statistical isotropy in the power spectrum and in the higher-order correlators. 

Based on this result, another way through which higher spin d.o.f can leave a distinct signatures in the inflationary fluctuations $\zeta$ has been studied in detail in Ref.~\cite{kehagias:2017cym}. In particular, the authors computed a general form of the anisotropic power spectra that arises from a generic spin-$s$ state. Its form can be written as in the following expansion (see later for more details):
\begin{eqnarray}
  \Braket{\zeta_{\vec{k}_1} \zeta_{\vec{k}_2}}
  &=& (2\pi)^3  \delta^{(3)}(\vec{k}_1 + \vec{k}_2) P_\zeta(k_1) \nonumber \\
  && \times \left[ 1 + \sum_{L \geq 1} \sum_{M} g_{LM} Y_{LM}(\hat{k}_1) \right]\, .
\label{eq:zeta2_gLM}
\end{eqnarray}
Here, the reality condition and parity invariance of the curvature power spectrum imposes $g_{LM} = (-1)^M g_{L, -M}^{*}$ and $g_{L= {\rm odd}, M} = 0$. This is the case we are going to focus on in this paper. One interesting feature is that, for a given state of spin $s$ the multipole coefficients $g_{LM}$ run up to $g_{(2s) M}$.  
A detection of these coefficients in the primordial power spectrum could reveal the presence of higher spin d.o.f during inflation. In particular, we have studied what the effect of these coefficients in the CMB and in the LSS galaxy power spectra is. Our main results are a forecast about the sensitivity of present and future CMB missions and LSS surveys to the multipole coefficients  $g_{LM}$ which set the level and the type of statistical anisotropy in the primordial power spectrum. 

Following Ref.~\cite{kehagias:2017cym} and to the best of our knowledge,  a prediction for nonvanishing coefficients $g_{L\geq 4, M}$ due to higher spin fields is derived for the first time in this paper where we also  provide a forecast on such coefficients.  As a concrete example we have focused, in particular, on the case of a spin $s=2$ field, but our result can be easily generalized to higher spin $s>2$. Indeed another interesting result that we find is the almost independence of our forecasts on $L$ and hence we expect that our results can be applied also to the case of a spinning state with spin $s>2$. In fact the case $s>2$ can be particularly relevant for models of inflation where higher spin fields are implemented consistently. Indeed, since the seminal work of Vasiliev \cite{Vas}, it is well known that massless higher spin field equations can be written in de Sitter spacetime, at the expense of introducing an infinite amount of spin states. As detailed below, we find that, e.g., by exploiting {\it Planck} data and present LSS surveys a sensitivity to $g_{4M}$ as low as $10^{-2}$ can be achieved, and that an order of magnitude improvement can be achieved for CMB and LSS ideal (cosmic-variance-limited) surveys. 

The paper is organized as follows. In Sec.~\ref{sec:summary} we briefly recall and summarize the main features of the models studied in Ref.~\cite{kehagias:2017cym}, starting from the expression for the statistical anisotropic power spectra, Eq.~\eqref{eq:zeta2_spin-s}, which is then expanded as in Eq.~\eqref{eq:zeta2_gLM}. In Sec.~\ref{sec:CMB} we derive the expressions for CMB angular power spectra induced by the anisotropic power spectra ~(\ref{eq:zeta2_spin-s}) and we derive our forecasts for CMB experiments. In Sec.~\ref{sec:LSS} we instead provide our results for the galaxy power spectra and the sensitivity of present and future LSS surveys to the imprint of higher spin states. Finally we comment on our results and we conclude in Sec.~\ref{sec:con}.

%%%%%%%%%%%%%%%%%%%%%%%%%%%%%%%%%%%%%%%%%%%%%%%%%%%%%%%%%%%%%%%%%%%%%%%%%%%%%%
\section{Anisotropic primordial curvature power spectra from higher spin fields}
\label{sec:summary}
%%%%%%%%%%%%%%%%%%%%%%%%%%%%%%%%%%%%%%%%%%%%%%%%%%%%%%%%%%%%%%%%%%%%%%%%%%%%%%
%\noindent

As we have mentioned in the Introduction, it is possible to generate anisotropic power spectra and bispectra in the primordial perturbation of the comoving curvature perturbation $\zeta$ by coupling  vector spin-1 perturbations to the inflaton field $\phi$ in such a way that they remain constant on super-Hubble scale. This is achieved by modifying the kinetic term of the vector field through a kinetic term of the form \cite{Ratra:1991bn,Martin:2007ue,Bartolo:2012sd}
\begin{equation}
{\cal L} = -\frac{1}{4}I(\phi)F^{\mu \nu} F_{\mu\nu} .
\label{L}
\end{equation}
Such a model has been first proposed as a magnetogenesis scenario in Ref.~\cite{Ratra:1991bn}, and it has been later considered in the context of anisotropic models of inflation sourced by a nonvanishing vacuum expectation value of the vector field. In the latter case, by suitably choosing the coupling function $I(\phi)$, it is possible to produce an almost constant vector energy density [and correspondingly nondecaying super-horizon (almost) scale invariant fluctuations of the vector field]. In fact one can introduce an ``electric field'' $E_i= - a^{-2} \langle I^{1/2} \rangle  A'_i$ , where $\langle \cdots \rangle$ stands for the vacuum expectation value, and a prime is a derivative with respect to the conformal time $d\tau=dt/a(t)$. Therefore, in the standard case, $I = \rm const.$, the classical equation of motion from Eq.~\eqref{L} gives $\bar{E}_i \propto 1/(a^2 \langle I^{1/2}\rangle) \propto a^{-2}$, while a constant ``electric'' field $\bar{E}_i$ (and hence a constant energy density $\rho_E = |\vec{E}|^2/2$) is generated with $\langle I^{1/2}\rangle \propto a^{-2}(\tau)$. Given a slow-roll potential for the inflaton field $\phi$, this can be achieved by arranging the functional form of $I(\phi)$ with the inflaton potential~\cite{Martin:2007ue}. In these models cosmological perturbations arising during inflation leave specific imprints in terms of a quadrupolar anisotropy in the primordial curvature power spectrum dictated by the vacuum expectation value of the nonvanishing background classical field. If, for instance, a classical background for the electric field $\bar{E}_i$ components is generated (that is for wavelengths much larger than the Hubble radius during inflation) along a given direction $\hat{p}$, then the two-point correlator of the curvature perturbation is modified as~\cite{Dulaney:2010sq,Gumrukcuoglu:2010yc,Watanabe:2010fh,Bartolo:2012sd,Biagetti:2013qqa}
\begin{equation}
  \Braket{\zeta_{\vec{k}_1} \zeta_{\vec{k}_2}}
  = (2\pi)^3 \delta^{(3)} (\vec{k}_1 + \vec{k}_2) \bar{P}_\zeta(k_1)\left( 1 + c_1\sin^{2}\theta_k \right), \label{eq:zeta2_spin-1}
\end{equation}
where $\bar{P}_\zeta$ is the isotropic part of the power spectrum, $\theta_k$ is the angle between the directions $\hat{p}$ and $\hat{k}_1$, and  the amplitude of the anisotropic modulation $c_1$  scales like $[{\bar E}^2/ (\epsilon H^2 M_{\rm pl}^2) ] N_k^2$, with $H$ the Hubble rate during inflation, $\epsilon=-\dot H/H^2$ one of the slow-roll parameters, $N_k$ the number of $e$-folds until the end of inflation calculated from the instant when the  wavelength $1/k$ leaves the Hubble radius, and $M_{\rm pl}$ the reduced Planck mass. The quadrupolar anisotropy arises because of the interactions between the curvature perturbation $\zeta$ and the electric vector field fluctuations $\delta E_i$. The Lagrangian~(\ref{L}) in fact gives rise to interaction terms of the type $\bar{E}_i \delta E_i \zeta$ and $(\delta E_i)^2 \zeta$, both generating a correction to the power spectrum due to the vector internal leg exchange among the external curvature $\zeta$ legs (see Ref.~\cite{Bartolo:2012sd} for the corresponding Feynman diagrams).

This example emphasizes the importance of the presence of spinning extra d.o.f during inflation. If minimally coupled to the spacetime background, massive higher spin fields modify  the squeezed limit of the non-Gaussian correlation functions of the curvature perturbation  when intermediate higher spin fields are exchanged in internal lines \cite{Arkani-Hamed:2015bza,Lee:2016vti,Meerburg:2016zdz}. The correction to the non-Gaussian correlators depends on their masses and  spins thus carrying information about these fundamental parameters. 
The fact that fields with spin $s$ may play a dynamical role only as virtual states is due to the fact that the   de Sitter isometries impose the so-called  Higuchi bound \cite{Higuchi:1986py} on their  masses 
\begin{equation}
m^2 > s(s-1)H^2.
\end{equation}
This implies that on super-Hubble scales the fluctuations of the higher spin fields decay at least as $e^{-Ht}$ \cite{Arkani-Hamed:2015bza} and their imprints onto the non-Gaussian correlators   are  suppressed by powers of the exchanged momentum in the squeezed configuration.

The example of the vector field teaches us however the lesson that spinning d.o.f can be long lived on super-Hubble scales if suitably coupled to the inflaton field. This has been recently investigated in Ref.~\cite{kehagias:2017cym} where, through  a bottom-up approach starting  from the equation of motion of the higher spin fields  and requiring the correct number of propagating d.o.f,  it has been shown that  there exist  couplings with the inflaton field which allow  the higher spin perturbations to remain  constant on scales larger than the Hubble radius. 

Inflation may offer therefore a unique chance to test the presence of spinning high-energy states. One possible way is the following. Similarly to the case of the vector field for which an infrared electric (or magnetic) component can be generated during inflation through the accumulation of the various perturbation modes exiting the Hubble radius before the 60 or so $e$-folds to the end of inflation~\cite{Bartolo:2012sd}, infrared modes of the higher spins can be generated. Indeed, even if a zero mode of the higher spin field is not present  at the beginning of inflation, it will be  generated with time with an amplitude of the order of the square root of its variance, $\langle \bar{\sigma}_{i_1\cdots i_s}\rangle\sim H^2 N$, with  $N$  the total number of $e$-folds~\cite{kehagias:2017cym}. Indeed, by properly coupling the higher spin field to a function $I(\phi)$, the helicity $s$ mode of such a field obeys the equation \cite{kehagias:2017cym} 
\begin{eqnarray}
&&\sigma_s''-\left[\frac{2(1-s)}{\tau}-I'\right] \sigma_s'
\nonumber \\
&&+\left[k^2+\frac{M_s^2/H^2+s^2-4s}{\tau^2}+\frac{s(1+\alpha)}{\tau}I'\right]
\sigma_s = 0,
\label{eqs}
\end{eqnarray}
where $M_s^2=(s+2)H^2(\alpha-s\alpha-1)/\alpha$,  $I(\phi(\tau))=\ln(-H\tau)/\alpha$ and $\alpha=1/[2(1-s)]$. For such values, the canonically normalized higher spin field $\bar{\sigma}_{s}={\rm exp}(I(\phi)/2)\sigma_s$ is quantum mechanically generated with a constant value on super-Hubble scales in the very same way the scalar perturbation $\zeta$ is.

This classical background breaks the isotropy.
Indeed,  if the higher spin field couples to the inflaton through a suitable interaction of the form
\begin{eqnarray}
S \supset g_s H^2 \int  {\rm d}^4 x \,e^{3H t}\,{\rm exp}(I(\phi))\sigma_{i_1\cdots i_s}\sigma^{i_1\cdots i_s},
\end{eqnarray}
with $g_s$ a spin dependent coupling, it leads to an anisotropic  correction to the comoving curvature power spectrum of the form \cite{kehagias:2017cym} 
\begin{equation} 
  \Braket{\zeta_{\vec{k}_1} \zeta_{\vec{k}_2}}
  %---
  = (2\pi)^3 \delta^{(3)} (\vec{k}_1 + \vec{k}_2)
  \bar{P}_\zeta(k_1)\left( 1 + c_s\sin^{2s}\theta_k \right), \label{eq:zeta2_spin-s}
\end{equation}
where $c_s$ scales as $[g_s^2\langle \bar{\sigma}\rangle^2/ (\epsilon H^2 M_{\rm pl}^2)] N_k^2$ and we have indicated by $\langle\bar{\sigma}\rangle$ the overall amplitude of the classical background $\langle \bar{\sigma}_{i_1\cdots i_s}\rangle$ and again with $\theta_k$ the angle between the directions $\hat{k}_1$ and $\hat{p}$, the latter identifying the special direction identified by $\langle \bar{\sigma}_{i_1\cdots i_s}\rangle$. 
In general we have that 
\begin{eqnarray}
 \langle \bar{\sigma}_{i_1\cdots i_s}\rangle=\langle \bar{\sigma}_s\rangle\,\Sigma_{i_1\cdots i_s},
 \end{eqnarray} 
 where 
 \begin{eqnarray}
   \Sigma_{i_1\cdots i_s} &=& {\rm Sym}\Big{[} \hat{p}_{i_1} \hat{p}_{i_2}\cdots \hat{p}_{i_s} + \delta_{i_1 i_2}\Sigma_{i_3 \cdots i_{s}} \Big{]} \nonumber \\
  && - {\rm Traces}\Big{[}\Sigma_{i_1\cdots i_s}\Big{]}, 
 \end{eqnarray}
 and ${\rm Sym}$ denotes complete symmetrization. For instance, for $s=2$, $s=3$, and $s=4$, we have
 \begin{align}
\langle \bar{\sigma}_{ij}\rangle &=\langle \bar{\sigma}_2\rangle\left(\hat{p}_i\hat{p}_j-\frac{1}{3}\delta_{ij}\right),\\
%---
\langle \bar{\sigma}_{ijk}\rangle &=\langle \bar{\sigma}_3\rangle
\left\{\hat{p}_i\hat{p}_j\hat{p}_k-\frac{1}{5}\left(\hat{p}_i \delta_{jk} + 2~{\rm perm}\right)\right\},\\
%---
\langle \bar{\sigma}_{ijkl}\rangle &=\langle \bar{\sigma}_4\rangle
\left\{ \hat{p}_i\hat{p}_j\hat{p}_k\hat{p}_l 
- \frac{1}{7}(\hat{p}_i \hat{p}_j \delta_{kl} + 5~{\rm perm})
\right.\nonumber \\
&\left.\qquad\qquad
+ \frac{1}{35}(\delta_{ij}\delta_{kl} + 2~{\rm perm})
\right\},
 \end{align}
 and so on. The goal of the paper is to investigate the capability of current and future cosmological observations to detect or put bounds on the anisotropic signatures induced in the power spectrum of the curvature perturbation.

%***************************************************************

For the following phenomenological analyses, we decompose the characteristic angular dependence in Eq.~\eqref{eq:zeta2_spin-s} into the spherical harmonic basis according to Eq.~\eqref{eq:zeta2_gLM}. As the amplitude parameter $c_s$ is constant in $\vec{k}_{1,2}$, we express $\sin^{2s}\theta_k$ in terms of the Legendre polynomials ${\cal L}_L(\cos\theta_k)$ as
\begin{eqnarray} 
  \sin^{2s}\theta_k &=& \sum_{L\geq 0} A_L {\cal L}_L(\cos\theta_k) \nonumber \\
  &=& A_0 + \sum_{L\geq 1} A_L {\cal L}_L(\cos\theta_k). 
\end{eqnarray}
In the last line the $L=0$ mode and the others are written separately. The Legendre coefficients are computed according to 
\begin{eqnarray}
  A_L &=& \frac{2L+1}{2}\int_0^\pi {\rm d} \theta_k \, \sin^{2s + 1}\theta_k \, {\cal L}_L(\cos\theta_k) \nonumber \\
  &=&
\frac{2L+1}{2} \int_{-1}^1 {\rm d}\mu_k \,  (1-\mu_k^2)^s \, {\cal L}_L(\mu_k), \label{eq:al}
\end{eqnarray}
where $\mu_k = \cos\theta_k = \hat k_1 \cdot \hat p$. Therefore, the comoving curvature power spectrum \eqref{eq:zeta2_spin-s} can be rewritten as  
\begin{eqnarray}
 \Braket{\zeta_{\vec{k}_1} \zeta_{\vec{k}_2}}
 &=& (2\pi)^3 \delta^{(3)}(\vec{k}_1 + \vec{k}_2) \bar{P}_\zeta(k)\left(1 + c_s A_0 \right) \nonumber \\
 && \times \left[ 1+\sum_{L\geq 1}\frac{c_s A_L}{1+c_s A_0}
 {\cal L}_{L} (\hat{k}_1 \cdot \hat{p})
 \right]. 
\end{eqnarray}
Comparing this with Eq.~\eqref{eq:zeta2_gLM} after the spherical harmonic decomposition:
\begin{eqnarray}
 {\cal L}_L(\hat{k}_1 \cdot \hat{p}) = \frac{4\pi}{2L + 1} \sum_M  Y_{LM}(\hat{k}_1) Y_{LM}^*(\hat{p}), \label{eq:Legendre}
\end{eqnarray}
we obtain $P_\zeta(k) = \bar{P}_\zeta(k)\left(1 + c_s A_0 \right)$ and 
\begin{equation}
g_{L M}=\frac{4\pi}{2L+1} \frac{c_s A_L}{1+c_s A_0} Y^*_{L M}(\hat p). 
\end{equation}
For $L \geq 0$ and $s \geq 0$, Eq.~\eqref{eq:al} is solved analytically as
\begin{equation}
A_L = \frac{2L+1}{2} \frac{\pi\,  \left[\Gamma(s+1)\right]^2}{\Gamma(\frac{1-L}{2})\Gamma(\frac{2+L}{2})\Gamma(\frac{2s-L+2}{2})\Gamma(\frac{2s+L+3}{2})},
\end{equation}
and we therefore find that
\begin{eqnarray}
  g_{L M} &=& \frac{2\pi^2\, c_s\,  \left[\Gamma(s+1)\right]^2}{\Gamma(\frac{1-L}{2})\Gamma(\frac{2+L}{2})\Gamma(\frac{2s-L+2}{2})\Gamma(\frac{2s+L+3}{2})} \nonumber \\ 
  &&\times  
  \frac{2\Gamma(\frac{2s+3}{2})}{2\Gamma(\frac{2s+3}{2})+c_s\sqrt{\pi}\Gamma(s+1)}  Y^*_{L M}(\hat p). \label{eq:glm}
\end{eqnarray}
It is easy to check that due to the poles of $\Gamma(z)$, we have $g_{L>2s, M} = g_{L = {\rm odd}, M} = 0$, so we see that there are $s$ nonvanishing multipole coefficients: $g_{2M}$, $g_{4M}$, $\cdots$, $g_{(2s-2)M}$ and $g_{(2s) M}$. Moreover, the contribution of large spins ($s\gg 1$) are not exponentially suppressed but rather
\begin{eqnarray}
g_{L M} \sim \frac{c_s}{\sqrt{s}}\,  \frac{2\pi^2}{\Gamma\left(\frac{1-L}{2}\right)\Gamma\left(\frac{2+L}{2}\right)} Y_{L M}^*(\hat{p}).
\end{eqnarray}

For example, nonzero $g_{2M}$ and $g_{4M}$ arise in the $s=2$ case and in this case
the curvature power spectrum is  
\begin{eqnarray}
  \Braket{\zeta_{\vec{k}_1} \zeta_{\vec{k}_2}}
  &=& (2\pi)^3  \delta^{(3)}(\vec{k}_1 + \vec{k}_2)P_\zeta(k_1) \nonumber \\ 
  && \times \left[ 1+\sum_{L=2,4} \sum_{M} g_{LM} Y_{LM}(\hat k _1) \right],  
\end{eqnarray}
where $P_\zeta(k) = \bar{P}_\zeta(k)[ 1 + (8 c_2 / 15) ] \approx  \bar{P}_\zeta(k)$ (assuming that $|c_2| \ll 1$), and $g_{2M}$ and $g_{4M}$ are given, according to Eq.~\eqref{eq:glm}, by
\begin{eqnarray}
  g_{2M} &=& - \frac{64 \pi c_2}{105 + 56 c_2} Y_{2M}^*(\hat{p})
  \approx - \frac{64\pi}{105} c_2  Y_{2M}^*(\hat{p})  , \label{eq:g2M_s2} \\
  g_{4M} &=& \frac{32\pi c_2}{315 + 168c_2} Y_{4M}^*(\hat{p})
  \approx \frac{32\pi}{315} c_2 Y_{4M}^*(\hat{p}). 
\end{eqnarray}

The quadrupolar coefficients $g_{2M}$ have been both theoretically and observationally well studied (see, e.g., Refs.~\cite{Ackerman:2007nb,Pullen:2007tu,Pullen:2010zy,Bartolo:2012sd,Ade:2013nlj,Ade:2015lrj,Ade:2015hxq,Shiraishi:2016omb,Shiraishi:2016wec,Dulaney:2010sq,Gumrukcuoglu:2010yc,Watanabe:2010fh,Bartolo:2012sd,Biagetti:2013qqa,Kim:2013gka,Bartolo:2014hwa,Naruko:2014bxa,Ashoorioon:2016lrg}. The latest limit $|g_{2M}| \lesssim 0.01$ \cite{Ade:2015lrj,Ade:2015hxq,Sugiyama:2017ggb}  (see also Ref.~\cite{Ade:2015ava}) can be directly translated into the bound on $c_2$ via Eq.~\eqref{eq:g2M_s2} if $\hat{p}$ is fixed or marginalized over. There are also the CMB constraints on higher-order coefficients \cite{Ramazanov:2013wea,Rubtsov:2014yua}, providing the information on $c_{s>2}$. From the next section, we analyze the induced signatures in the CMB and galaxy power spectra, and then focus especially on the impacts of the hexadecapolar term $g_{4M}$.

%%%%%%%%%%%%%%%%%%%%%%%%%%%%%%%%%%%%%%%%%%%%%%%%%%%%%%%%%%%%%%%%%%%%%%%%%%%%%%
\section{Anisotropic signatures in the CMB power spectra}
\label{sec:CMB}
%%%%%%%%%%%%%%%%%%%%%%%%%%%%%%%%%%%%%%%%%%%%%%%%%%%%%%%%%%%%%%%%%%%%%%%%%%%%%%

\begin{figure*}[t]
\begin{tabular}{cc}
   \begin{minipage}{0.5\hsize}
     \begin{center}
       \includegraphics[width=1\textwidth]{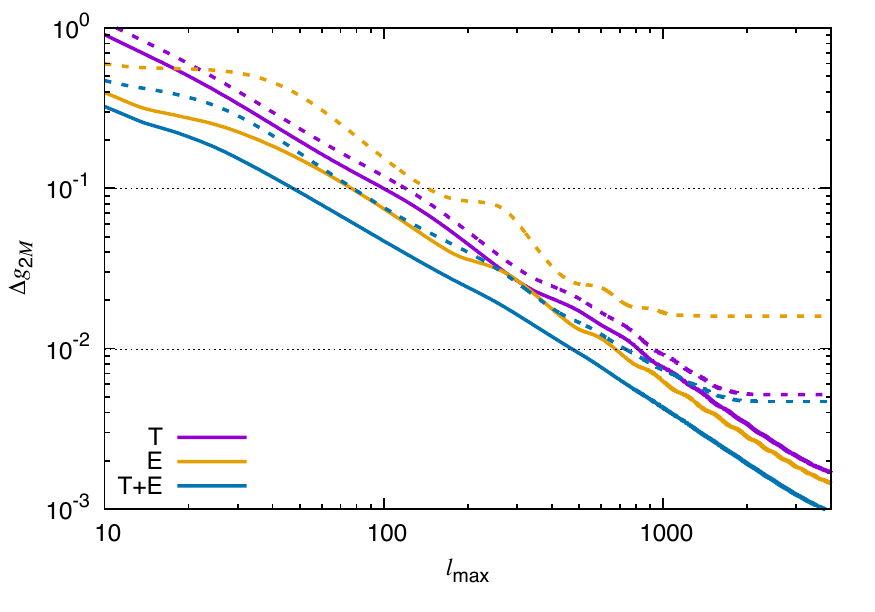}
     \end{center}
   \end{minipage}
   \begin{minipage}{0.5\hsize}
     \begin{center}
       \includegraphics[width=1\textwidth]{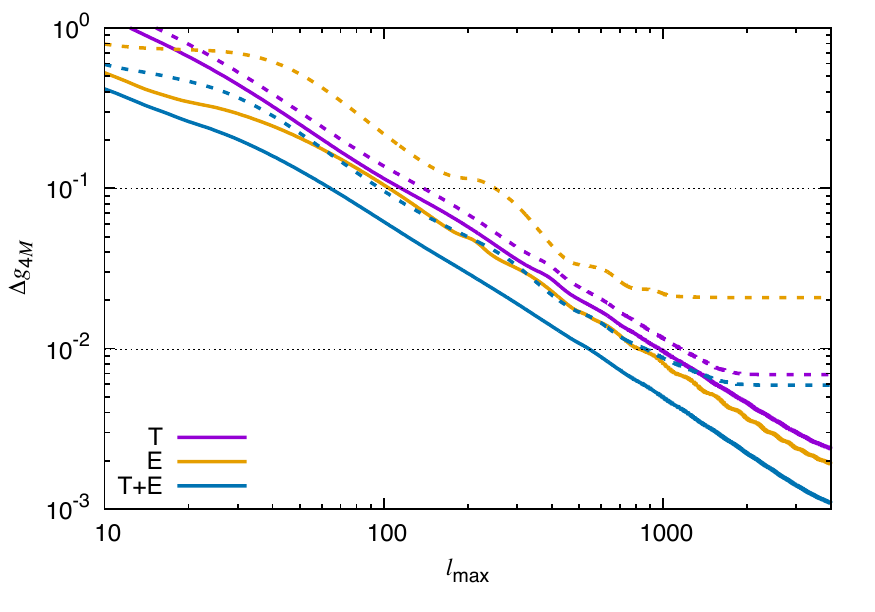}
     \end{center}
   \end{minipage}
\end{tabular}
\caption{Expected $1\sigma$ errors on $g_{2M}$ (left panel) and $g_{4M}$ (right panel) computed from temperature alone (purple lines), E-mode polarization alone (yellow lines), and temperature and E-mode polarization jointly (blue lines), by assuming a CMB noiseless CVL-level survey with $f_{\rm sky} = 1$ (solid lines) and a {\it Planck}-like one with $f_{\rm sky} = 0.7$ (dashed lines). We here take $\ell_{\rm min} = 2$.}
\label{fig:error_CMB} 
\end{figure*}

At linear order, the harmonic coefficients of the CMB temperature ($X = T$) and E-mode polarization ($X = E$) anisotropies sourced by the curvature perturbation are expressed as
\begin{eqnarray}
  a_{\ell m}^X 
  &=& 4\pi i^\ell 
  \int \frac{{\rm d}^3 k}{(2\pi)^3} Y_{\ell m}^*(\hat{k}) \zeta_{\vec{k}} 
 {\cal T}_{\ell}^X(k) , \label{eq:alm_iso}
\end{eqnarray}
where ${\cal T}_{\ell}^X(k)$ is the scalar-mode CMB transfer function. The angular power spectra induced by Eq.~\eqref{eq:zeta2_gLM} are derived from this, reading
\begin{eqnarray}
  \Braket{\prod_{n=1}^2 a_{\ell_n m_n}^{X_n}}
  &=& G_{\ell_1 \ell_2}^{X_1 X_2} (-1)^{m_1} \delta_{\ell_1, \ell_2} \delta_{m_1, -m_2} \nonumber \\
  && + (-1)^{m_2} C_{\ell_1 m_1, \ell_2 -m_2}^{X_1 X_2} ,
\end{eqnarray}
where
\begin{eqnarray}
  G_{\ell_1 \ell_2}^{X_1 X_2}  &=& \frac{2}{\pi} \int_0^\infty k^2 {\rm d} k P_\zeta(k)
       {\cal T}_{\ell_1}^{X_1}(k)
       {\cal T}_{\ell_2}^{X_2}(k) ,
         \label{eq:Gfunc} \\
      %-------
  C_{\ell_1 m_1, \ell_2 m_2}^{X_1 X_2} &=& i^{\ell_1 -\ell_2}   G_{\ell_1 \ell_2}^{X_1 X_2}
  (-1)^{m_1} \sum_{L \geq 1} h_{\ell_1 \ell_2 L} \nonumber \\
  && \times 
   \sum_{M} g_{LM} 
\left(
  \begin{array}{ccc}
  \ell_1 & \ell_2 & L \\
   -m_1 & m_2 & M 
  \end{array}
  \right) ,
   \label{eq:Cl1l2_gLM}
\end{eqnarray}
with 
\begin{equation}
h_{l_1 l_2 l_3} \equiv \sqrt{\frac{(2 l_1 + 1)(2 l_2 + 1)(2 l_3 + 1)}{4 \pi}}
\left(
  \begin{array}{ccc}
  l_1 & l_2 & l_3 \\
  0 & 0 & 0
  \end{array}
  \right).
  \end{equation}
We notice that the nonzero $g_{LM}$'s generate not only diagonal ($\ell_1 = \ell_2$) but also off-diagonal ($\ell_1 \neq \ell_2$) modes in $C_{\ell_1 m_1, \ell_2 m_2}^{X_1 X_2}$ \eqref{eq:Cl1l2_gLM}. To be specific, nonvanishing modes actually rely on the selection rules of $h_{\ell_1 \ell_2 L}$; namely, $|\ell_2 - L| \leq \ell_1 \leq |\ell_2 + L|$ and $\ell_1 + \ell_2 + L = {\rm even}$. If $g_{4M}$ is nonzero, the modes obeying $|\ell_1 - \ell_2| = 2, 4$ do not vanish.%
\footnote{These off-diagonal components may also be induced by galactic foregrounds \cite{Kamionkowski:2014wza}.}
In the same manner, $g_{(2s)M}$ induces the signal in $|\ell_1 - \ell_2| = 2, 4, \cdots, 2(s-1), 2s$. This is indeed due to the fact that statistical isotropy is broken in these models.

  We are now interested in how accurately $g_{LM}$ could be extracted from (future) CMB data. For this goal, we therefore perform a Fisher matrix analysis. Under the diagonal covariance matrix approximation, the Fisher matrix computed from the temperature and E-mode polarization anisotropies is given as \cite{Hanson:2009gu, Hanson:2010gu,Ma:2011ii}
\begin{eqnarray}
  F_{LM, L' M'}^{T+E}
&=& \frac{f_{\rm sky}}{2} 
  \sum_{\ell_1 m_1 \ell_2 m_2}
  \sum_{\substack{X_1 X_2 \\ X_1' X_2'}} 
  \frac{\partial C_{\ell_1 m_1 \ell_2 m_2}^{X_1 X_2}}{\partial g_{L M}^{*}}  \nonumber \\
  && \times (C^{-1})_{\ell_1}^{X_1 X_1'}  (C^{-1})_{\ell_2}^{X_2 X_2'} 
  \frac{\partial C_{\ell_1 m_1 \ell_2 m_2}^{X_1' X_2' *}}{\partial g_{L'M'}}, 
  \end{eqnarray}
where $f_{\rm sky}$ is the fraction of the sky coverage and $C^{-1}$ is the inverse of the $2 \times 2$ power spectrum matrix:
\begin{eqnarray}
  C_\ell^{XX'}
  = \left(
  \begin{array}{ccc}
    G_{\ell \ell}^{TT} + N_\ell^{TT} & G_{\ell \ell}^{TE}  \\
    G_{\ell \ell}^{TE} & G_{\ell \ell}^{EE} + N_\ell^{EE}  
  \end{array}
  \right) \, ,
\end{eqnarray}
with $N_\ell^{XX}$ denoting the noise spectra of the temperature and E-mode polarization. Plugging Eq.~\eqref{eq:Cl1l2_gLM} into this leads to
\begin{eqnarray}
  F_{LM, L' M'}^{T+E}
  &=& \frac{f_{\rm sky}}{2} \frac{\delta_{L, L'} \delta_{M, M'}}{2L+1}
  \sum_{\ell_1, \ell_2 = \ell_{\rm min}}^{\ell_{\rm max}} h_{\ell_1 \ell_2 L}^2
  \sum_{\substack{X_1 X_2 \\ X_1' X_2'}} G_{\ell_1 \ell_2}^{X_1 X_2}
  \nonumber \\ 
&& \times
(C^{-1})_{\ell_1}^{X_1 X_1'}
(C^{-1})_{\ell_2}^{X_2 X_2'} 
G_{\ell_1 \ell_2}^{X_1' X_2'}. \label{eq:FishCMB3D_gLM}
  \end{eqnarray}
The 1$\sigma$ errors can be computed by $\Delta g_{LM}^{T+E} = 1 / \sqrt{F_{LM, LM}^{T+E}}$.

The Fisher matrix coming from the temperature or E-mode autocorrelation is given by a subset of this matrix, reading
\begin{equation}
 F_{LM, L' M'}^X =
  \frac{f_{\rm sky}}{2} \frac{\delta_{LL'} \delta_{MM'}}{2L+1}
  \sum_{\ell_1, \ell_2 = \ell_{\rm min}}^{\ell_{\rm max}} h_{\ell_1 \ell_2 L}^2 
      \frac{\left(G_{\ell_1 \ell_2}^{X X}\right)^2}{C_{\ell_1}^{XX} C_{\ell_2}^{XX}},
\end{equation}
where the 1$\sigma$ errors read $\Delta g_{LM}^X = 1/\sqrt{F_{LM, L M}^X}$. In a noiseless cosmic-variance-limited (CVL) measurement (i.e. $C_\ell^{XX} \simeq G_{\ell \ell}^{XX}$), $\left(G_{\ell_1 \ell_2}^{X X}\right)^2 / [C_{\ell_1}^{XX} C_{\ell_2}^{XX}] \simeq 1$ is justified, so the Fisher matrix can be simplified to 
\begin{equation}
F_{LM, LM}^{X ({\rm CVL})} \simeq \frac{f_{\rm sky}}{2(2L+1)}
\sum_{\ell_1, \ell_2 = \ell_{\rm min}}^{\ell_{\rm max}} h_{\ell_1 \ell_2 L}^2 
\simeq \frac{f_{\rm sky}}{8\pi} \ell_{\rm max}^2 ,
\end{equation}
where we have dropped the subdominant contributions from $\ell_{\rm min}$ by assuming $\ell_{\rm max} \gg \ell_{\rm min}$. This yields
\begin{eqnarray}
  \Delta g_{LM}^{X ({\rm CVL})} \simeq \sqrt{\frac{8\pi}{f_{\rm sky}}} \ell_{\rm max}^{-1} \, .\label{eq:dgLM_CMB_CV}
\end{eqnarray}
Notice that the latter result indicates a very weak dependence of $\Delta g_{LM}$ on $L$.

Figure~\ref{fig:error_CMB} describes the numerical results of $\Delta g_{2M}$ and $\Delta g_{4M}$ estimated from temperature/polarization alone and temperature and polarization jointly, as a function of $\ell_{\rm max}$. Note that our results of $\Delta g_{2M}$ are in agreement with those obtained in the previous literature \cite{Pullen:2007tu,Ma:2011ii}. We find there that, in a full-sky noiseless CVL measurement, $g_{4M} = {\cal O}(10^{-3})$ is detectable if $\ell_{\rm max} \gtrsim 1000$. This is consistent with an expectation from Eq.~\eqref{eq:dgLM_CMB_CV}. Figure~\ref{fig:error_CMB} also includes the errors expected in a {\it Planck}-like realistic survey. To compute these, non-negligible $N_\ell^{TT}$ and $N_\ell^{EE}$ close to the {\it Planck} noise level \cite{Planck:2006aa,Ade:2015xua} and $f_{\rm sky} = 0.7$ are taken into account. These induce sensitivity reduction, while $\Delta g_{4M}^{T}$ and $\Delta g_{4M}^{T+E}$ can still go below $10^{-2}$ for $\ell_{\rm max} \gtrsim 1000$.

It is also confirmed from this figure that $\Delta g_{4M}$ has a size substantially similar to $\Delta g_{2M}$. This supports an expectation from Eq.~\eqref{eq:dgLM_CMB_CV} that $\Delta g_{LM}$ is almost independent of $L$; hence, one can expect $\Delta g_{(2s)M} = {\cal O}(10^{-3})$ for $\ell_{\rm max} \gtrsim 1000$.

%%%%%%%%%%%%%%%%%%%%%%%%%%%%%%%%%%%%%%%%%%%%%%%%%%%%%%%%%%%%%%%%%%%%%%%%%%%%%%
\section{Anisotropic signatures in the galaxy power spectra}
\label{sec:LSS}
%%%%%%%%%%%%%%%%%%%%%%%%%%%%%%%%%%%%%%%%%%%%%%%%%%%%%%%%%%%%%%%%%%%%%%%%%%%%%%

In this section, we discuss the search for $g_{LM}$ with the galaxy power spectrum. The primordial curvature power spectrum under consideration is statistically anisotropic but homogeneous, so the resulting redshift-space galaxy power spectrum can be written as
\begin{eqnarray}
  \Braket{\delta^s(\vec{k}_1) \delta^s(\vec{k}_2) }
  = (2\pi)^3 \delta^{(3)}(\vec{k}_1 + \vec{k}_2 ) P^s(\vec{k}_1, \hat{n}) ,
\end{eqnarray}
with $\hat{n}$ being a line-of-sight direction. Here, we have used the local plane parallel approximation that is justified when the visual angle for correlation scales of interest is small. We therefore ignore the wide-angle effect in the following analysis. For simplicity, an argument of time, redshift $z$, is here and hereinafter omitted in some variables.

Let us work under a scenario that isotropy of the Universe is broken during inflation, while it is restored after that and large-scale density fluctuations grow linearly. We can then model the galaxy power spectrum according to \cite{Kaiser:1987qv,Hamilton:1997zq}
\begin{eqnarray}
  P^s(\vec{k}, \hat{n}) = P_m(\vec{k})
  \left[b + f (\hat{k} \cdot \hat{n})^2 \right]^2 , 
\end{eqnarray}
where $b(z)$ is the linear bias parameter, and $f(z) \equiv \partial \ln D / \partial \ln a$ with $a$ and $D(a)$ denoting the scale factor and the growth factor, respectively. The quadrupolar angular dependence in the second term comes from the redshift-space distortion (RSD). In our case, the matter power spectrum is given by
\begin{eqnarray}
  P_m(\vec{k}) = M_k^2 P_\zeta(k) 
  \left[ 1 + \sum_{L\geq 1}\sum_{M} g_{LM} Y_{LM}(\hat{k}) \right] , \label{eq:deltam2_g2M}
\end{eqnarray}
where $M_k(z)$ is the linear matter transfer function. Decomposing the contribution $M_k^2 P_\zeta(k)\left[b + f (\hat{k} \cdot \hat{n})^2 \right]^2$ into the Legendre basis, the galaxy power spectrum from Eq.~\eqref{eq:zeta2_gLM} is rewritten as 
\begin{eqnarray}
  P^s(\vec{k}, \hat{n}) &=& \left[ \sum_{j} P_j(k) {\cal L}_j(\hat{k} \cdot \hat{n}) \right] \nonumber \\
  && \times \left[ 1 + \sum_{L\geq 1}\sum_{M} g_{LM} Y_{LM}(\hat{k}) \right] , \label{eq:deltag2}
\end{eqnarray} 
with%
\footnote{The power spectrum in usual isotropic universe models can always be expanded as in the first line of Eq.~\eqref{eq:deltag2}. We only consider RSD here for simplicity but the contribution of the other relativistic effects (i.e., Doppler, Sachs-Wolfe effect, etc.) is simply a correction on the coefficients $P_j$, with $j = 0, \cdots ,4$. The only exception is lensing as its contribution is still in the form of Eq.~\eqref{eq:deltag2} but affects all multipoles and it is not restricted to $j \leq 4$ (see Refs.~\cite{Raccanelli:2013gja,Tansella:2017rpi}).}
\begin{eqnarray}
 P_0(k) &=& \left(b^2 + \frac{2}{3}b f + \frac{1}{5} f^2 \right) M_k^2 P_\zeta(k)   , \\
 P_2(k) &=& \left( \frac{4}{3}b f  + \frac{4}{7} f^2 \right) M_k^2 P_\zeta(k), \\
 P_4(k) &=& \frac{8}{35} f^2 M_k^2 P_\zeta(k), \\
 P_1(k) &=& P_3(k) = P_{j \geq 5}(k) = 0.
\end{eqnarray}

%----------------------

\begin{figure*}[t]
 \begin{tabular}{cc}
   \begin{minipage}{0.5\hsize}
     \begin{center}
       \includegraphics[width=1\textwidth]{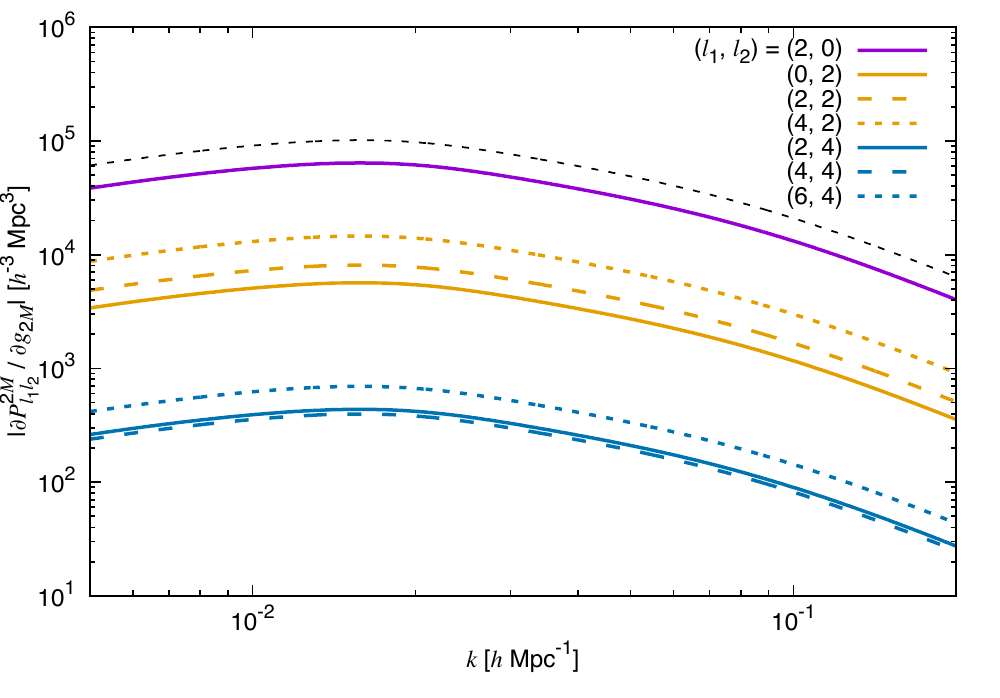}
  \end{center}
   \end{minipage}
   \begin{minipage}{0.5\hsize}
     \begin{center}
       \includegraphics[width=1\textwidth]{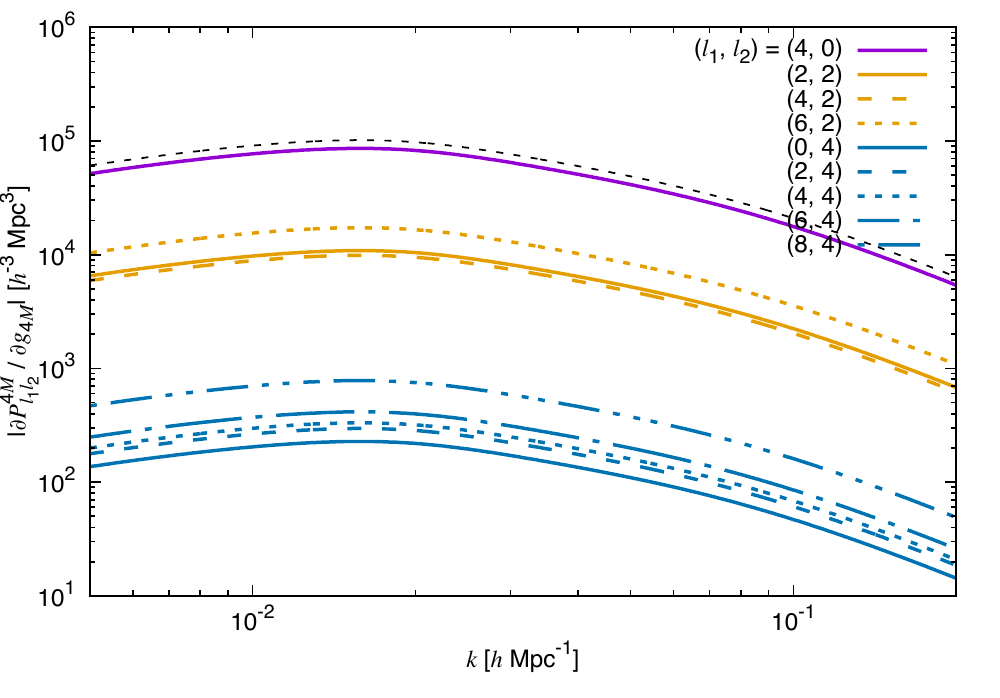}
  \end{center}
   \end{minipage}
 \end{tabular}
 \caption{All nonvanishing BipoSH coefficients for $L=2$ and $4$: $\partial P_{l_1 l_2}^{2M} / \partial g_{2M}$ and $\partial P_{l_1 l_2}^{4M} / \partial g_{4M}$. The black dashed lines describe $P_0$. We here take $b = 2.0$ and $z = 0.5$. The lines for $L = 2$ are fully consistent with those in Ref.~\cite{Shiraishi:2016wec}.}
  \label{fig:Pl1l2} 
\end{figure*}

%-------

Reference~\cite{Shiraishi:2016wec} found that the peculiar angular dependence in the anisotropic curvature power spectrum can be completely distinguished from the RSD one via the bipolar spherical harmonic (BipoSH) decomposition \cite{Varshalovich:1988ye,Szalay:1997cc,Hajian:2003qq}:
\begin{eqnarray}
 P^s(\vec{k}, \hat{n}) = \sum_{\ell \ell' LM} \pi_{\ell \ell'}^{LM}(k) X_{\ell \ell'}^{LM}(\hat{k},\hat{n}) ,
\end{eqnarray}
where the BipoSH basis is \cite{Varshalovich:1988ye}
\begin{eqnarray}
  X_{\ell \ell'}^{LM}(\hat{k},\hat{n})
  &\equiv& 
  \sum_{mm'} {\cal C}_{\ell m \ell' m'}^{LM} Y_{\ell m}(\hat{k}) Y_{\ell' m'}(\hat{n})
  \label{eq:S_basis}
\end{eqnarray}
with
\begin{equation}
  {\cal C}_{l_1 m_1 l_2 m_2}^{l_3 m_3} \equiv (-1)^{l_1 - l_2 + m_3} \sqrt{2l_3 + 1} \left( \begin{array}{ccc} l_1 & l_2 & l_3 \\ m_1 & m_2 & -m_3 \end{array}  \right)
\end{equation}
  denoting the Clebsch-Gordan coefficients.%
\footnote{Even if the wide-angle effect that is ignored in this paper is taken into account, the primordial anisotropic signal can be cleanly extracted by means of the tripolar spherical harmonic decomposition \cite{Shiraishi:2016wec}.}
The BipoSH coefficients are derived according to
\begin{eqnarray}
\pi_{\ell\ell'}^{LM}(k) =  \int {\rm d}^2 \hat{k} \int {\rm d}^2 \hat{n} P^s(\vec{k}, \hat{n}) X_{\ell \ell'}^{LM *}(\hat{k},\hat{n}).
\end{eqnarray}
The angular dependence in $P^s(\vec{k}, \hat{n})$ \eqref{eq:deltag2} is decomposed using the spherical harmonics as Eq.~\eqref{eq:Legendre}. In the same manner as Ref.~\cite{Shiraishi:2016wec}, performing the angular integrals of the spherical harmonics and adding the induced angular momenta, we can simplify $\pi_{\ell\ell'}^{LM}$ from Eq.~\eqref{eq:deltag2}. Renormalizing it as
\begin{equation}
  P_{\ell \ell'}^{L M} \equiv \pi_{\ell \ell'}^{LM} (-1)^L \sqrt{\frac{(2L+1)(2\ell+1)(2\ell'+1)}{(4\pi)^2 } } H_{\ell \ell' L}
\end{equation}
without loss of generality, we derive \cite{Sugiyama:2017ggb} 
\begin{equation}
    P_{\ell \ell'}^{L M}(k) =
    \begin{cases}
      P_{\ell}(k) \delta_{\ell, \ell'}  \delta_{M,0} & (L = 0) \\
      P_{\ell'}(k) \sqrt{\frac{2L+1}{4\pi } } (2\ell+1) H_{\ell \ell' L}^2  g_{LM}  & (L \geq 1)
    \end{cases},
  \end{equation}
where $H_{l_1 l_2 l_3} \equiv \left(
  \begin{array}{ccc}
  l_1 & l_2 & l_3 \\
  0 & 0 & 0
  \end{array}
  \right)$. It is obvious from this equation that the $L \neq 0$ mode of $P_{\ell \ell'}^{LM}$ becomes an unbiased estimator for $g_{LM}$. Nonvanishing combinations of multipoles in the $L \neq 0$ mode are determined by the triangular inequality and parity-even condition of $H_{\ell \ell' L}$ (e.g., $P_{20}^{2M}$, $P_{02}^{2M}$, $P_{22}^{2M}$, $P_{42}^{2M}$, $P_{24}^{2M}$, $P_{44}^{2M}$ and $P_{64}^{2M}$ for $L=2$, or $P_{40}^{4M}$, $P_{22}^{4M}$, $P_{42}^{4M}$, $P_{62}^{4M}$, $P_{04}^{4M}$, $P_{24}^{4M}$, $P_{44}^{4M}$, $P_{64}^{4M}$, and $P_{84}^{4M}$ for $L = 4$).

All nonvanishing components of 
\begin{eqnarray}
  \frac{\partial P_{l_1 l_2}^{L M}(k) }{\partial g_{L M}}
  = P_{l_2}(k) 
 \sqrt{\frac{2L+1}{4\pi}} (2 l_1 + 1) H_{l_1 l_2 L}^2 
\end{eqnarray}
for $L=2$ and $4$ are plotted in Fig.~\ref{fig:Pl1l2}. It is confirmed that the dominant signal lies in $\ell' = 0$ (i.e., $P_{20}^{2M}$ for $L=2$ or $P_{40}^{4M}$ for $L = 4$). This is simply because of $P_0 \gg P_2 \gg P_4$.

%-------
\begin{figure*}[t]
\begin{tabular}{cc}
   \begin{minipage}{0.5\hsize}
     \begin{center}
   \includegraphics[width=1\textwidth]{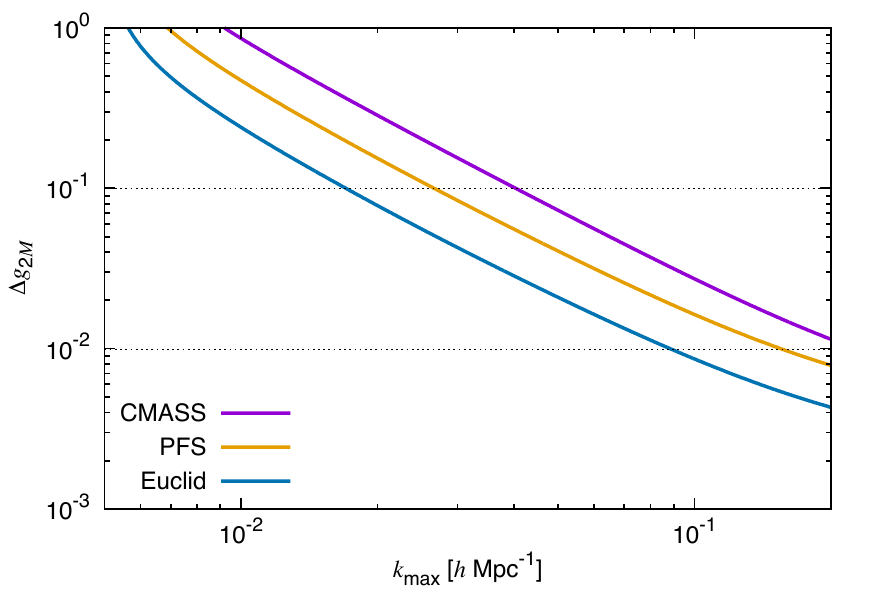}
  \end{center}
   \end{minipage}
   \begin{minipage}{0.5\hsize}
     \begin{center}
    \includegraphics[width=1\textwidth]{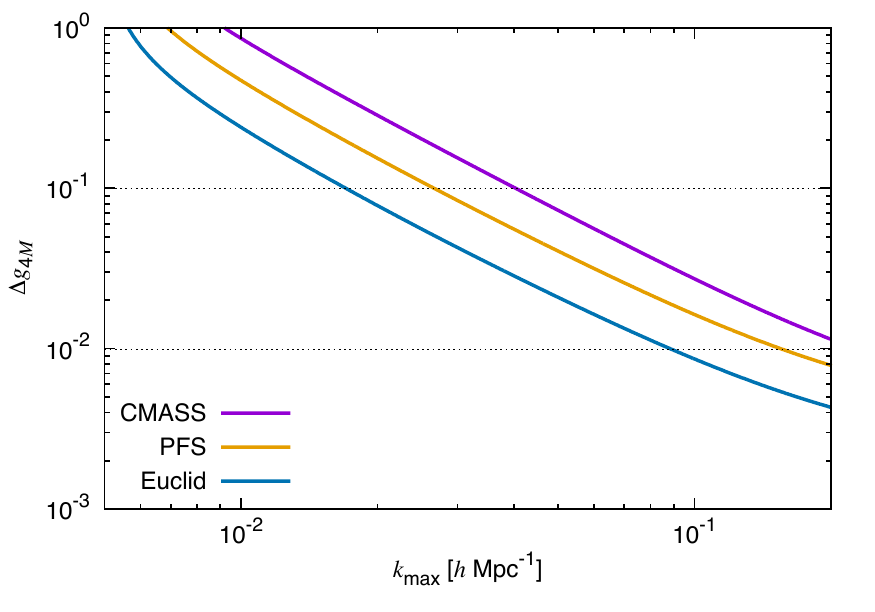}
  \end{center}
   \end{minipage}
 \end{tabular}
\caption{Expected $1\sigma$ errors $\Delta g_{2M}$ and $\Delta g_{4M}$ in CMASS, PFS and Euclid. For PFS and Euclid, the co-add information from multi-redshift slices is taken into account via Eq.~\eqref{eq:Fish_tomography} with an assumption that different redshift bins are uncorrelated. The results of $\Delta g_{2M}$ are consistent with those in Ref.~\cite{Shiraishi:2016wec}. Here we take $k_{\rm min} = 0.005 h ~{\rm Mpc}^{-1}$}
  \label{fig:error_gal} 
\end{figure*}
%----------------

Via a diagonal covariance matrix approximation, the Fisher matrix from $P_{\ell \ell'}^{L M}(k)$ is simplified to \cite{Shiraishi:2016wec}
\begin{eqnarray}
  F_{LM, L' M'}^g &=& \delta_{L, L'} \delta_{M, M'} V \int_{k_{\rm min}}^{k_{\rm max}} \frac{k^2 {\rm d}k}{ 2\pi^2} \sum_{l_1 l_2 l_1' l_2'} \frac{\partial P_{l_1 l_2}^{L M}(k) }{\partial g_{L M}} \nonumber \\
  && \times \left(\Theta^{-1}\right)_{l_1 l_2, l_1' l_2'}^{L}(k)
  \left( \frac{\partial P_{l_1' l_2'}^{L M}(k) }{\partial g_{L M}} \right)^* , \label{eq:Fishgal}
\end{eqnarray}
where $V$ is the survey volume and
\begin{eqnarray}
  \Theta_{l_1 l_2, l_1' l_2'}^L(k) 
  &=& (2l_1 + 1)(2l_2 + 1)(2l_1' + 1)(2l_2' + 1) \nonumber \\
 &\times& 
 (2L+1)  (-1)^{l_1} \left[1 + (-1)^{l_1'}\right]
  H_{l_1 l_2 L} H_{l_1' l_2' L}  \nonumber \\
 &\times& \sum_{J J'} P_J^{(\rm O)}(k) P_{J'}^{(\rm O)}(k)
   \sum_{L_1 L_2}  
   (2L_1 + 1)  \nonumber \\ 
  &\times&
  (2L_2 + 1) H_{l_1 J L_1} H_{l_2 J L_2} H_{l_1' J' L_1} H_{l_2' J' L_2} \nonumber \\ 
  &\times&
  \left\{
      \begin{array}{ccc}
      L & L_1 & L_2\\
      J & l_2 & l_1 
      \end{array}
    \right\}
    \left\{
    \begin{array}{ccc}
      L & L_1 & L_2\\
      J' & l_2' & l_1' 
    \end{array}
    \right\} . \label{eq:Theta}
\end{eqnarray}
The Legendre coefficients $ P_J^{(\rm O)}$ are composed of cosmic variance and the homogeneous shot noise according to $ P_0^{(\rm O)} = P_0 + 1 / n_g$, $P_{2}^{(\rm O)} = P_{2}$, $P_{4}^{(\rm O)} = P_{4}$, and $P_{1}^{(\rm O)} = P_{3}^{(\rm O)} = P_{J \geq 5}^{(\rm O)} = 0$ with $n_g$ denoting the number density of galaxies. 

As indicated above, the signal for $l_2 = l_2' = 0$ contributes dominantly to the summation in Eq.~\eqref{eq:Fishgal}. This, together with $H_{l_1 0 L} \propto \delta_{l_1, L}$, allows us to write
\begin{eqnarray}
  F_{LM, LM}^g &\simeq&  V \int_{k_{\rm min}}^{k_{\rm max}} \frac{k^2 {\rm d}k}{ 8\pi^3}  
  \frac{(2L+1) \left[ P_{0}(k) \right]^2 }{\Theta_{L 0, L 0}^{L}(k)} .
\end{eqnarray}
The fact that $P_0^{\rm (O)} \gg P_{J \geq 1}^{\rm (O)}$ reduces the covariance matrix to 
\begin{equation}
  \Theta_{L 0, L 0}^L(k) 
  \simeq 
  \begin{cases}
    2 (2L + 1) \left[P_0^{(\rm O)}(k)\right]^2  & (L = {\rm even}) \\
    0 & (L = {\rm odd})
  \end{cases}.
\end{equation}
Assuming a CVL-level galaxy survey (i.e., $P_0^{(\rm O)} \simeq P_0$), the $k$ integral can be analytically performed and we thus obtain
\begin{eqnarray}
  F_{LM, LM}^{g ({\rm CVL})} &\simeq&  \frac{V}{48\pi^3 } 
      \left(k_{\rm max}^3 - k_{\rm min}^3 \right) . \label{eq:Fishgal_CV}
\end{eqnarray}
Again this indicates the weak dependence of $\Delta g_{LM}$ on $L$.

Unlike the CMB power spectrum, the galaxy one has redshift dependence. This enables a tomographic analysis. Adding the information from $N_{\rm bin}$ independent redshift bins, the Fisher matrix is enhanced as
\begin{eqnarray}
  F_{LM, LM}^{g, \rm tot} = \sum_{i=1}^{N_{\rm bin}} F_{LM, LM}^g(z_i) . \label{eq:Fish_tomography}
\end{eqnarray}
The expected $1\sigma$ errors on $ g_{LM}$ are computed according to $\Delta g_{LM}^g = 1 / \sqrt{F_{LM,LM}^{g, \rm tot}}$.%
\footnote{
  Here, we consider the information from equal-time galaxy correlators for simplicity, while adding that from different-time ones will improve the sensitivity.
}

Figure~\ref{fig:error_gal} shows $\Delta g_{2M}$ and $\Delta g_{4M}$ expected in an ongoing survey like the Baryon Oscillation Spectroscopic Survey \cite{Bolton:2012hz,Dawson:2012va} that is part of SDSS-III \cite{Eisenstein:2011sa} (CMASS) and next generation ones like the Subaru Prime Focus Spectrograph (PFS) \cite{Ellis:2012rn}, and Euclid \cite{Laureijs:2011gra}, as a function of $k_{\rm max}$. The values of $b$, $n_g$, and $V$ per each redshift bin in each experiment adopted here are summarized in Ref.~\cite{Shiraishi:2016wec}. In comparison with  $\Delta g_{2M}$ evaluated from the previous SDSS data \cite{Pullen:2010zy}, our results shrink by $1-2$ orders of magnitude because of the sensitivity improvement. One can confirm that $\Delta g_{2M}$ and $\Delta g_{4M}$ scale like $k_{\rm max}^{-3/2}$ as expected from Eq.~\eqref{eq:Fishgal_CV}. The difference of their overall size between each experiment is mainly due to the difference of the total survey volume. The results in Fig.~\ref{fig:error_gal} and the weak dependence of $\Delta g_{LM}$ on $L$ suggest that $g_{(2s)M} = {\cal O}(10^{-2})$ is already testable from currently available data, and $g_{(2s)M} = {\cal O}(10^{-3})$ could be measured in near future.

%%%%%%%%%%%%%%%%%%%%%%%%%%%%%%%%%%%%%%%%%%%%%%%%%%%%%%%%%%%%%%%%%%%%%%%%%%%%%%
\section{Conclusions}\label{sec:con}

In this paper we have taken the first step towards the detection of possible signatures of higher spin fields during inflation in the specific model where these fields are rendered effectively massless by a suitable coupling to the inflaton field, making them  long living on super-Hubble scales. We have shown that in this setup these higher spin fields may leave a distinct  feature at the level of the power spectrum of the comoving curvature perturbation by generating statistical anisotropy parametrized by the coefficients $ g_{LM}$ in the spherical harmonic decomposition of $P_\zeta$. 

We have shown that these coefficients can be probed down to $ {\cal O}(10^{-3})$ through CMB and LSS experiments, to $ {\cal O}(10^{-2})$ in current surveys and ${\cal O}(10^{-3})$ in a near future survey. A remarkable feature is that the forecasted errors on the $ g_{LM}$ are nearly independent on $L$. This is certainly welcome as one should expect that,  in a consistent theory of higher spin fields in a de Sitter phase, all the spins, if effectively massless, should play a role. It would be interesting to investigate the imprint of higher spin fields
on higher-order correlators and observables.

%%%%%%%%%%%%%%%%%%%%%%%%%%%%%%%%%%%%%%%%%%%%%%%%%%%%%%%%%%%%%%%%%%%%%%%%%%%%%

\acknowledgements

M.\,S. is supported by a JSPS Grant-in-Aid for Research Activity Start-up Grant No.~17H07319. M.\,S. also acknowledges the Center for Computational Astrophysics, National Astronomical Observatory of Japan, for providing the computing resources of Cray XC30. A.\,R. is supported by the Swiss National Science Foundation (SNSF), project {\sl Investigating the Nature of Dark Matter}, Project No.~200020-159223. V.\,T.  acknowledges support by the SNSF. N.\,B. and M.\,L. acknowledge financial support by ASI Grant No.~2016-24-H.0. N.\,B. and M.\,L. acknowledge partial financial support by the ASI/INAF Agreement I/072/09/0 for the Planck LFI Activity of Phase E2.

%%%%%%%%%%%%%%%%%%%%%%%%%%%%%%%%%%%%%%%%%%%%%%%%%%%%%%%%%%%%%%%%%%%%%%%%%%%%%%%%%%%%%%%%%%%%%%%%%%%%%%%%%%%%%%%%%%%%%%%%%%%%%%%%%%%%%%%%%%%%%%%%%%%%%%%%%%%%%%%%%%%%%%%%%%%%%%
% \appendix

%########################################
% Create the reference section using BibTeX
\bibliography{paper}

\begin{thebibliography}{74}%
\makeatletter
\providecommand \@ifxundefined [1]{%
 \@ifx{#1\undefined}
}%
\providecommand \@ifnum [1]{%
 \ifnum #1\expandafter \@firstoftwo
 \else \expandafter \@secondoftwo
 \fi
}%
\providecommand \@ifx [1]{%
 \ifx #1\expandafter \@firstoftwo
 \else \expandafter \@secondoftwo
 \fi
}%
\providecommand \natexlab [1]{#1}%
\providecommand \enquote  [1]{``#1''}%
\providecommand \bibnamefont  [1]{#1}%
\providecommand \bibfnamefont [1]{#1}%
\providecommand \citenamefont [1]{#1}%
\providecommand \href@noop [0]{\@secondoftwo}%
\providecommand \href [0]{\begingroup \@sanitize@url \@href}%
\providecommand \@href[1]{\@@startlink{#1}\@@href}%
\providecommand \@@href[1]{\endgroup#1\@@endlink}%
\providecommand \@sanitize@url [0]{\catcode `\\12\catcode `\$12\catcode
  `\&12\catcode `\#12\catcode `\^12\catcode `\_12\catcode `\%12\relax}%
\providecommand \@@startlink[1]{}%
\providecommand \@@endlink[0]{}%
\providecommand \url  [0]{\begingroup\@sanitize@url \@url }%
\providecommand \@url [1]{\endgroup\@href {#1}{\urlprefix }}%
\providecommand \urlprefix  [0]{URL }%
\providecommand \Eprint [0]{\href }%
\providecommand \doibase [0]{http://dx.doi.org/}%
\providecommand \selectlanguage [0]{\@gobble}%
\providecommand \bibinfo  [0]{\@secondoftwo}%
\providecommand \bibfield  [0]{\@secondoftwo}%
\providecommand \translation [1]{[#1]}%
\providecommand \BibitemOpen [0]{}%
\providecommand \bibitemStop [0]{}%
\providecommand \bibitemNoStop [0]{.\EOS\space}%
\providecommand \EOS [0]{\spacefactor3000\relax}%
\providecommand \BibitemShut  [1]{\csname bibitem#1\endcsname}%
\let\auto@bib@innerbib\@empty
%</preamble>
\bibitem [{\citenamefont {Ade}\ \emph {et~al.}(2016{\natexlab{a}})\citenamefont
  {Ade} \emph {et~al.}}]{Ade:2015lrj}%
  \BibitemOpen
  \bibfield  {author} {\bibinfo {author} {\bibfnamefont {P.~A.~R.}\
  \bibnamefont {Ade}} \emph {et~al.} (\bibinfo {collaboration} {Planck}),\
  }\href {\doibase 10.1051/0004-6361/201525898} {\bibfield  {journal} {\bibinfo
   {journal} {Astron. Astrophys.}\ }\textbf {\bibinfo {volume} {594}},\
  \bibinfo {pages} {A20} (\bibinfo {year} {2016}{\natexlab{a}})},\ \Eprint
  {http://arxiv.org/abs/1502.02114} {arXiv:1502.02114 [astro-ph.CO]}
  \BibitemShut {NoStop}%
%%CITATION = ARXIV:1502.02114;%%
\bibitem [{\citenamefont {Ade}\ \emph {et~al.}(2016{\natexlab{b}})\citenamefont
  {Ade} \emph {et~al.}}]{Ade:2015ava}%
  \BibitemOpen
  \bibfield  {author} {\bibinfo {author} {\bibfnamefont {P.~A.~R.}\
  \bibnamefont {Ade}} \emph {et~al.} (\bibinfo {collaboration} {Planck}),\
  }\href {\doibase 10.1051/0004-6361/201525836} {\bibfield  {journal} {\bibinfo
   {journal} {Astron. Astrophys.}\ }\textbf {\bibinfo {volume} {594}},\
  \bibinfo {pages} {A17} (\bibinfo {year} {2016}{\natexlab{b}})},\ \Eprint
  {http://arxiv.org/abs/1502.01592} {arXiv:1502.01592 [astro-ph.CO]}
  \BibitemShut {NoStop}%
%%CITATION = ARXIV:1502.01592;%%
\bibitem [{\citenamefont {Ade}\ \emph {et~al.}(2016{\natexlab{c}})\citenamefont
  {Ade} \emph {et~al.}}]{Ade:2015xua}%
  \BibitemOpen
  \bibfield  {author} {\bibinfo {author} {\bibfnamefont {P.~A.~R.}\
  \bibnamefont {Ade}} \emph {et~al.} (\bibinfo {collaboration} {Planck}),\
  }\href {\doibase 10.1051/0004-6361/201525830} {\bibfield  {journal} {\bibinfo
   {journal} {Astron. Astrophys.}\ }\textbf {\bibinfo {volume} {594}},\
  \bibinfo {pages} {A13} (\bibinfo {year} {2016}{\natexlab{c}})},\ \Eprint
  {http://arxiv.org/abs/1502.01589} {arXiv:1502.01589 [astro-ph.CO]}
  \BibitemShut {NoStop}%
%%CITATION = ARXIV:1502.01589;%%
\bibitem [{\citenamefont {Bartolo}\ \emph {et~al.}(2004)\citenamefont
  {Bartolo}, \citenamefont {Komatsu}, \citenamefont {Matarrese},\ and\
  \citenamefont {Riotto}}]{Bartolo:2004if}%
  \BibitemOpen
  \bibfield  {author} {\bibinfo {author} {\bibfnamefont {N.}~\bibnamefont
  {Bartolo}}, \bibinfo {author} {\bibfnamefont {E.}~\bibnamefont {Komatsu}},
  \bibinfo {author} {\bibfnamefont {S.}~\bibnamefont {Matarrese}}, \ and\
  \bibinfo {author} {\bibfnamefont {A.}~\bibnamefont {Riotto}},\ }\href
  {\doibase 10.1016/j.physrep.2004.08.022} {\bibfield  {journal} {\bibinfo
  {journal} {Phys. Rept.}\ }\textbf {\bibinfo {volume} {402}},\ \bibinfo
  {pages} {103} (\bibinfo {year} {2004})},\ \Eprint
  {http://arxiv.org/abs/astro-ph/0406398} {arXiv:astro-ph/0406398 [astro-ph]}
  \BibitemShut {NoStop}%
%%CITATION = ASTRO-PH/0406398;%%
\bibitem [{\citenamefont {Chen}(2010)}]{Chen:2010xka}%
  \BibitemOpen
  \bibfield  {author} {\bibinfo {author} {\bibfnamefont {X.}~\bibnamefont
  {Chen}},\ }\href {\doibase 10.1155/2010/638979} {\bibfield  {journal}
  {\bibinfo  {journal} {Adv. Astron.}\ }\textbf {\bibinfo {volume} {2010}},\
  \bibinfo {pages} {638979} (\bibinfo {year} {2010})},\ \Eprint
  {http://arxiv.org/abs/1002.1416} {arXiv:1002.1416 [astro-ph.CO]} \BibitemShut
  {NoStop}%
%%CITATION = ARXIV:1002.1416;%%
\bibitem [{\citenamefont {Byrnes}\ and\ \citenamefont
  {Choi}(2010)}]{Byrnes:2010em}%
  \BibitemOpen
  \bibfield  {author} {\bibinfo {author} {\bibfnamefont {C.~T.}\ \bibnamefont
  {Byrnes}}\ and\ \bibinfo {author} {\bibfnamefont {K.-Y.}\ \bibnamefont
  {Choi}},\ }\href {\doibase 10.1155/2010/724525} {\bibfield  {journal}
  {\bibinfo  {journal} {Adv. Astron.}\ }\textbf {\bibinfo {volume} {2010}},\
  \bibinfo {pages} {724525} (\bibinfo {year} {2010})},\ \Eprint
  {http://arxiv.org/abs/1002.3110} {arXiv:1002.3110 [astro-ph.CO]} \BibitemShut
  {NoStop}%
%%CITATION = ARXIV:1002.3110;%%
\bibitem [{\citenamefont {Chen}\ and\ \citenamefont
  {Wang}(2010{\natexlab{a}})}]{Chen:2009we}%
  \BibitemOpen
  \bibfield  {author} {\bibinfo {author} {\bibfnamefont {X.}~\bibnamefont
  {Chen}}\ and\ \bibinfo {author} {\bibfnamefont {Y.}~\bibnamefont {Wang}},\
  }\href {\doibase 10.1103/PhysRevD.81.063511} {\bibfield  {journal} {\bibinfo
  {journal} {Phys. Rev.}\ }\textbf {\bibinfo {volume} {D81}},\ \bibinfo {pages}
  {063511} (\bibinfo {year} {2010}{\natexlab{a}})},\ \Eprint
  {http://arxiv.org/abs/0909.0496} {arXiv:0909.0496 [astro-ph.CO]} \BibitemShut
  {NoStop}%
%%CITATION = ARXIV:0909.0496;%%
\bibitem [{\citenamefont {Chen}\ and\ \citenamefont
  {Wang}(2010{\natexlab{b}})}]{Chen:2009zp}%
  \BibitemOpen
  \bibfield  {author} {\bibinfo {author} {\bibfnamefont {X.}~\bibnamefont
  {Chen}}\ and\ \bibinfo {author} {\bibfnamefont {Y.}~\bibnamefont {Wang}},\
  }\href {\doibase 10.1088/1475-7516/2010/04/027} {\bibfield  {journal}
  {\bibinfo  {journal} {JCAP}\ }\textbf {\bibinfo {volume} {1004}},\ \bibinfo
  {pages} {027} (\bibinfo {year} {2010}{\natexlab{b}})},\ \Eprint
  {http://arxiv.org/abs/0911.3380} {arXiv:0911.3380 [hep-th]} \BibitemShut
  {NoStop}%
%%CITATION = ARXIV:0911.3380;%%
\bibitem [{\citenamefont {Dimastrogiovanni}\ \emph {et~al.}(2016)\citenamefont
  {Dimastrogiovanni}, \citenamefont {Fasiello},\ and\ \citenamefont
  {Kamionkowski}}]{Dimastrogiovanni:2015pla}%
  \BibitemOpen
  \bibfield  {author} {\bibinfo {author} {\bibfnamefont {E.}~\bibnamefont
  {Dimastrogiovanni}}, \bibinfo {author} {\bibfnamefont {M.}~\bibnamefont
  {Fasiello}}, \ and\ \bibinfo {author} {\bibfnamefont {M.}~\bibnamefont
  {Kamionkowski}},\ }\href {\doibase 10.1088/1475-7516/2016/02/017} {\bibfield
  {journal} {\bibinfo  {journal} {JCAP}\ }\textbf {\bibinfo {volume} {1602}},\
  \bibinfo {pages} {017} (\bibinfo {year} {2016})},\ \Eprint
  {http://arxiv.org/abs/1504.05993} {arXiv:1504.05993 [astro-ph.CO]}
  \BibitemShut {NoStop}%
%%CITATION = ARXIV:1504.05993;%%
\bibitem [{\citenamefont {Baumann}\ and\ \citenamefont
  {Green}(2012)}]{Baumann:2011nk}%
  \BibitemOpen
  \bibfield  {author} {\bibinfo {author} {\bibfnamefont {D.}~\bibnamefont
  {Baumann}}\ and\ \bibinfo {author} {\bibfnamefont {D.}~\bibnamefont
  {Green}},\ }\href {\doibase 10.1103/PhysRevD.85.103520} {\bibfield  {journal}
  {\bibinfo  {journal} {Phys. Rev.}\ }\textbf {\bibinfo {volume} {D85}},\
  \bibinfo {pages} {103520} (\bibinfo {year} {2012})},\ \Eprint
  {http://arxiv.org/abs/1109.0292} {arXiv:1109.0292 [hep-th]} \BibitemShut
  {NoStop}%
%%CITATION = ARXIV:1109.0292;%%
\bibitem [{\citenamefont {Noumi}\ \emph {et~al.}(2013)\citenamefont {Noumi},
  \citenamefont {Yamaguchi},\ and\ \citenamefont {Yokoyama}}]{Noumi:2012vr}%
  \BibitemOpen
  \bibfield  {author} {\bibinfo {author} {\bibfnamefont {T.}~\bibnamefont
  {Noumi}}, \bibinfo {author} {\bibfnamefont {M.}~\bibnamefont {Yamaguchi}}, \
  and\ \bibinfo {author} {\bibfnamefont {D.}~\bibnamefont {Yokoyama}},\ }\href
  {\doibase 10.1007/JHEP06(2013)051} {\bibfield  {journal} {\bibinfo  {journal}
  {JHEP}\ }\textbf {\bibinfo {volume} {06}},\ \bibinfo {pages} {051} (\bibinfo
  {year} {2013})},\ \Eprint {http://arxiv.org/abs/1211.1624} {arXiv:1211.1624
  [hep-th]} \BibitemShut {NoStop}%
%%CITATION = ARXIV:1211.1624;%%
\bibitem [{\citenamefont {Kehagias}\ and\ \citenamefont
  {Riotto}(2015)}]{Kehagias:2015jha}%
  \BibitemOpen
  \bibfield  {author} {\bibinfo {author} {\bibfnamefont {A.}~\bibnamefont
  {Kehagias}}\ and\ \bibinfo {author} {\bibfnamefont {A.}~\bibnamefont
  {Riotto}},\ }\href {\doibase 10.1002/prop.201500025} {\bibfield  {journal}
  {\bibinfo  {journal} {Fortsch. Phys.}\ }\textbf {\bibinfo {volume} {63}},\
  \bibinfo {pages} {531} (\bibinfo {year} {2015})},\ \Eprint
  {http://arxiv.org/abs/1501.03515} {arXiv:1501.03515 [hep-th]} \BibitemShut
  {NoStop}%
%%CITATION = ARXIV:1501.03515;%%
\bibitem [{\citenamefont {Barnaby}\ \emph {et~al.}(2012)\citenamefont
  {Barnaby}, \citenamefont {Namba},\ and\ \citenamefont
  {Peloso}}]{Barnaby:2012tk}%
  \BibitemOpen
  \bibfield  {author} {\bibinfo {author} {\bibfnamefont {N.}~\bibnamefont
  {Barnaby}}, \bibinfo {author} {\bibfnamefont {R.}~\bibnamefont {Namba}}, \
  and\ \bibinfo {author} {\bibfnamefont {M.}~\bibnamefont {Peloso}},\ }\href
  {\doibase 10.1103/PhysRevD.85.123523} {\bibfield  {journal} {\bibinfo
  {journal} {Phys. Rev.}\ }\textbf {\bibinfo {volume} {D85}},\ \bibinfo {pages}
  {123523} (\bibinfo {year} {2012})},\ \Eprint {http://arxiv.org/abs/1202.1469}
  {arXiv:1202.1469 [astro-ph.CO]} \BibitemShut {NoStop}%
%%CITATION = ARXIV:1202.1469;%%
\bibitem [{\citenamefont {Bartolo}\ \emph {et~al.}(2013)\citenamefont
  {Bartolo}, \citenamefont {Matarrese}, \citenamefont {Peloso},\ and\
  \citenamefont {Ricciardone}}]{Bartolo:2012sd}%
  \BibitemOpen
  \bibfield  {author} {\bibinfo {author} {\bibfnamefont {N.}~\bibnamefont
  {Bartolo}}, \bibinfo {author} {\bibfnamefont {S.}~\bibnamefont {Matarrese}},
  \bibinfo {author} {\bibfnamefont {M.}~\bibnamefont {Peloso}}, \ and\ \bibinfo
  {author} {\bibfnamefont {A.}~\bibnamefont {Ricciardone}},\ }\href {\doibase
  10.1103/PhysRevD.87.023504} {\bibfield  {journal} {\bibinfo  {journal} {Phys.
  Rev.}\ }\textbf {\bibinfo {volume} {D87}},\ \bibinfo {pages} {023504}
  (\bibinfo {year} {2013})},\ \Eprint {http://arxiv.org/abs/1210.3257}
  {arXiv:1210.3257 [astro-ph.CO]} \BibitemShut {NoStop}%
%%CITATION = ARXIV:1210.3257;%%
\bibitem [{\citenamefont {Shiraishi}\ \emph {et~al.}(2013)\citenamefont
  {Shiraishi}, \citenamefont {Komatsu}, \citenamefont {Peloso},\ and\
  \citenamefont {Barnaby}}]{Shiraishi:2013vja}%
  \BibitemOpen
  \bibfield  {author} {\bibinfo {author} {\bibfnamefont {M.}~\bibnamefont
  {Shiraishi}}, \bibinfo {author} {\bibfnamefont {E.}~\bibnamefont {Komatsu}},
  \bibinfo {author} {\bibfnamefont {M.}~\bibnamefont {Peloso}}, \ and\ \bibinfo
  {author} {\bibfnamefont {N.}~\bibnamefont {Barnaby}},\ }\href {\doibase
  10.1088/1475-7516/2013/05/002} {\bibfield  {journal} {\bibinfo  {journal}
  {JCAP}\ }\textbf {\bibinfo {volume} {1305}},\ \bibinfo {pages} {002}
  (\bibinfo {year} {2013})},\ \Eprint {http://arxiv.org/abs/1302.3056}
  {arXiv:1302.3056 [astro-ph.CO]} \BibitemShut {NoStop}%
%%CITATION = ARXIV:1302.3056;%%
\bibitem [{\citenamefont {Bartolo}\ \emph
  {et~al.}(2015{\natexlab{a}})\citenamefont {Bartolo}, \citenamefont
  {Matarrese}, \citenamefont {Peloso},\ and\ \citenamefont
  {Shiraishi}}]{Bartolo:2015dga}%
  \BibitemOpen
  \bibfield  {author} {\bibinfo {author} {\bibfnamefont {N.}~\bibnamefont
  {Bartolo}}, \bibinfo {author} {\bibfnamefont {S.}~\bibnamefont {Matarrese}},
  \bibinfo {author} {\bibfnamefont {M.}~\bibnamefont {Peloso}}, \ and\ \bibinfo
  {author} {\bibfnamefont {M.}~\bibnamefont {Shiraishi}},\ }\href {\doibase
  10.1088/1475-7516/2015/07/039} {\bibfield  {journal} {\bibinfo  {journal}
  {JCAP}\ }\textbf {\bibinfo {volume} {1507}},\ \bibinfo {pages} {039}
  (\bibinfo {year} {2015}{\natexlab{a}})},\ \Eprint
  {http://arxiv.org/abs/1505.02193} {arXiv:1505.02193 [astro-ph.CO]}
  \BibitemShut {NoStop}%
%%CITATION = ARXIV:1505.02193;%%
\bibitem [{\citenamefont {Endlich}\ \emph {et~al.}(2013)\citenamefont
  {Endlich}, \citenamefont {Nicolis},\ and\ \citenamefont
  {Wang}}]{Endlich:2012pz}%
  \BibitemOpen
  \bibfield  {author} {\bibinfo {author} {\bibfnamefont {S.}~\bibnamefont
  {Endlich}}, \bibinfo {author} {\bibfnamefont {A.}~\bibnamefont {Nicolis}}, \
  and\ \bibinfo {author} {\bibfnamefont {J.}~\bibnamefont {Wang}},\ }\href
  {\doibase 10.1088/1475-7516/2013/10/011} {\bibfield  {journal} {\bibinfo
  {journal} {JCAP}\ }\textbf {\bibinfo {volume} {1310}},\ \bibinfo {pages}
  {011} (\bibinfo {year} {2013})},\ \Eprint {http://arxiv.org/abs/1210.0569}
  {arXiv:1210.0569 [hep-th]} \BibitemShut {NoStop}%
%%CITATION = ARXIV:1210.0569;%%
\bibitem [{\citenamefont {Shiraishi}\ \emph {et~al.}(2012)\citenamefont
  {Shiraishi}, \citenamefont {Nitta}, \citenamefont {Yokoyama},\ and\
  \citenamefont {Ichiki}}]{Shiraishi:2012rm}%
  \BibitemOpen
  \bibfield  {author} {\bibinfo {author} {\bibfnamefont {M.}~\bibnamefont
  {Shiraishi}}, \bibinfo {author} {\bibfnamefont {D.}~\bibnamefont {Nitta}},
  \bibinfo {author} {\bibfnamefont {S.}~\bibnamefont {Yokoyama}}, \ and\
  \bibinfo {author} {\bibfnamefont {K.}~\bibnamefont {Ichiki}},\ }\href
  {\doibase 10.1088/1475-7516/2012/03/041} {\bibfield  {journal} {\bibinfo
  {journal} {JCAP}\ }\textbf {\bibinfo {volume} {1203}},\ \bibinfo {pages}
  {041} (\bibinfo {year} {2012})},\ \Eprint {http://arxiv.org/abs/1201.0376}
  {arXiv:1201.0376 [astro-ph.CO]} \BibitemShut {NoStop}%
%%CITATION = ARXIV:1201.0376;%%
\bibitem [{\citenamefont {Shiraishi}(2012)}]{Shiraishi:2012sn}%
  \BibitemOpen
  \bibfield  {author} {\bibinfo {author} {\bibfnamefont {M.}~\bibnamefont
  {Shiraishi}},\ }\href {\doibase 10.1088/1475-7516/2012/06/015} {\bibfield
  {journal} {\bibinfo  {journal} {JCAP}\ }\textbf {\bibinfo {volume} {1206}},\
  \bibinfo {pages} {015} (\bibinfo {year} {2012})},\ \Eprint
  {http://arxiv.org/abs/1202.2847} {arXiv:1202.2847 [astro-ph.CO]} \BibitemShut
  {NoStop}%
%%CITATION = ARXIV:1202.2847;%%
\bibitem [{\citenamefont {Arkani-Hamed}\ and\ \citenamefont
  {Maldacena}(2015)}]{Arkani-Hamed:2015bza}%
  \BibitemOpen
  \bibfield  {author} {\bibinfo {author} {\bibfnamefont {N.}~\bibnamefont
  {Arkani-Hamed}}\ and\ \bibinfo {author} {\bibfnamefont {J.}~\bibnamefont
  {Maldacena}},\ }\href@noop {} {\  (\bibinfo {year} {2015})},\ \Eprint
  {http://arxiv.org/abs/1503.08043} {arXiv:1503.08043 [hep-th]} \BibitemShut
  {NoStop}%
%%CITATION = ARXIV:1503.08043;%%
\bibitem [{\citenamefont {Lee}\ \emph {et~al.}(2016)\citenamefont {Lee},
  \citenamefont {Baumann},\ and\ \citenamefont {Pimentel}}]{Lee:2016vti}%
  \BibitemOpen
  \bibfield  {author} {\bibinfo {author} {\bibfnamefont {H.}~\bibnamefont
  {Lee}}, \bibinfo {author} {\bibfnamefont {D.}~\bibnamefont {Baumann}}, \ and\
  \bibinfo {author} {\bibfnamefont {G.~L.}\ \bibnamefont {Pimentel}},\ }\href
  {\doibase 10.1007/JHEP12(2016)040} {\bibfield  {journal} {\bibinfo  {journal}
  {JHEP}\ }\textbf {\bibinfo {volume} {12}},\ \bibinfo {pages} {040} (\bibinfo
  {year} {2016})},\ \Eprint {http://arxiv.org/abs/1607.03735} {arXiv:1607.03735
  [hep-th]} \BibitemShut {NoStop}%
%%CITATION = ARXIV:1607.03735;%%
\bibitem [{\citenamefont {Biagetti}\ \emph {et~al.}(2017)\citenamefont
  {Biagetti}, \citenamefont {Dimastrogiovanni},\ and\ \citenamefont
  {Fasiello}}]{Biagetti:2017viz}%
  \BibitemOpen
  \bibfield  {author} {\bibinfo {author} {\bibfnamefont {M.}~\bibnamefont
  {Biagetti}}, \bibinfo {author} {\bibfnamefont {E.}~\bibnamefont
  {Dimastrogiovanni}}, \ and\ \bibinfo {author} {\bibfnamefont
  {M.}~\bibnamefont {Fasiello}},\ }\href {\doibase
  10.1088/1475-7516/2017/10/038} {\bibfield  {journal} {\bibinfo  {journal}
  {JCAP}\ }\textbf {\bibinfo {volume} {1710}},\ \bibinfo {pages} {038}
  (\bibinfo {year} {2017})},\ \Eprint {http://arxiv.org/abs/1708.01587}
  {arXiv:1708.01587 [astro-ph.CO]} \BibitemShut {NoStop}%
%%CITATION = ARXIV:1708.01587;%%
\bibitem [{\citenamefont {Sefusatti}\ \emph {et~al.}(2012)\citenamefont
  {Sefusatti}, \citenamefont {Fergusson}, \citenamefont {Chen},\ and\
  \citenamefont {Shellard}}]{Sefusatti:2012ye}%
  \BibitemOpen
  \bibfield  {author} {\bibinfo {author} {\bibfnamefont {E.}~\bibnamefont
  {Sefusatti}}, \bibinfo {author} {\bibfnamefont {J.~R.}\ \bibnamefont
  {Fergusson}}, \bibinfo {author} {\bibfnamefont {X.}~\bibnamefont {Chen}}, \
  and\ \bibinfo {author} {\bibfnamefont {E.~P.~S.}\ \bibnamefont {Shellard}},\
  }\href {\doibase 10.1088/1475-7516/2012/08/033} {\bibfield  {journal}
  {\bibinfo  {journal} {JCAP}\ }\textbf {\bibinfo {volume} {1208}},\ \bibinfo
  {pages} {033} (\bibinfo {year} {2012})},\ \Eprint
  {http://arxiv.org/abs/1204.6318} {arXiv:1204.6318 [astro-ph.CO]} \BibitemShut
  {NoStop}%
%%CITATION = ARXIV:1204.6318;%%
\bibitem [{\citenamefont {Norena}\ \emph {et~al.}(2012)\citenamefont {Norena},
  \citenamefont {Verde}, \citenamefont {Barenboim},\ and\ \citenamefont
  {Bosch}}]{Norena:2012yi}%
  \BibitemOpen
  \bibfield  {author} {\bibinfo {author} {\bibfnamefont {J.}~\bibnamefont
  {Norena}}, \bibinfo {author} {\bibfnamefont {L.}~\bibnamefont {Verde}},
  \bibinfo {author} {\bibfnamefont {G.}~\bibnamefont {Barenboim}}, \ and\
  \bibinfo {author} {\bibfnamefont {C.}~\bibnamefont {Bosch}},\ }\href
  {\doibase 10.1088/1475-7516/2012/08/019} {\bibfield  {journal} {\bibinfo
  {journal} {JCAP}\ }\textbf {\bibinfo {volume} {1208}},\ \bibinfo {pages}
  {019} (\bibinfo {year} {2012})},\ \Eprint {http://arxiv.org/abs/1204.6324}
  {arXiv:1204.6324 [astro-ph.CO]} \BibitemShut {NoStop}%
%%CITATION = ARXIV:1204.6324;%%
\bibitem [{\citenamefont {Meerburg}\ \emph {et~al.}(2017)\citenamefont
  {Meerburg}, \citenamefont {Münchmeyer}, \citenamefont {Muñoz},\ and\
  \citenamefont {Chen}}]{Meerburg:2016zdz}%
  \BibitemOpen
  \bibfield  {author} {\bibinfo {author} {\bibfnamefont {P.~D.}\ \bibnamefont
  {Meerburg}}, \bibinfo {author} {\bibfnamefont {M.}~\bibnamefont
  {Münchmeyer}}, \bibinfo {author} {\bibfnamefont {J.~B.}\ \bibnamefont
  {Muñoz}}, \ and\ \bibinfo {author} {\bibfnamefont {X.}~\bibnamefont {Chen}},\
  }\href {\doibase 10.1088/1475-7516/2017/03/050} {\bibfield  {journal}
  {\bibinfo  {journal} {JCAP}\ }\textbf {\bibinfo {volume} {1703}},\ \bibinfo
  {pages} {050} (\bibinfo {year} {2017})},\ \Eprint
  {http://arxiv.org/abs/1610.06559} {arXiv:1610.06559 [astro-ph.CO]}
  \BibitemShut {NoStop}%
%%CITATION = ARXIV:1610.06559;%%
\bibitem [{\citenamefont {Chen}\ \emph
  {et~al.}(2016{\natexlab{a}})\citenamefont {Chen}, \citenamefont {Dvorkin},
  \citenamefont {Huang}, \citenamefont {Namjoo},\ and\ \citenamefont
  {Verde}}]{Chen:2016vvw}%
  \BibitemOpen
  \bibfield  {author} {\bibinfo {author} {\bibfnamefont {X.}~\bibnamefont
  {Chen}}, \bibinfo {author} {\bibfnamefont {C.}~\bibnamefont {Dvorkin}},
  \bibinfo {author} {\bibfnamefont {Z.}~\bibnamefont {Huang}}, \bibinfo
  {author} {\bibfnamefont {M.~H.}\ \bibnamefont {Namjoo}}, \ and\ \bibinfo
  {author} {\bibfnamefont {L.}~\bibnamefont {Verde}},\ }\href {\doibase
  10.1088/1475-7516/2016/11/014} {\bibfield  {journal} {\bibinfo  {journal}
  {JCAP}\ }\textbf {\bibinfo {volume} {1611}},\ \bibinfo {pages} {014}
  (\bibinfo {year} {2016}{\natexlab{a}})},\ \Eprint
  {http://arxiv.org/abs/1605.09365} {arXiv:1605.09365 [astro-ph.CO]}
  \BibitemShut {NoStop}%
%%CITATION = ARXIV:1605.09365;%%
\bibitem [{\citenamefont {Ballardini}\ \emph {et~al.}(2016)\citenamefont
  {Ballardini}, \citenamefont {Finelli}, \citenamefont {Fedeli},\ and\
  \citenamefont {Moscardini}}]{Ballardini:2016hpi}%
  \BibitemOpen
  \bibfield  {author} {\bibinfo {author} {\bibfnamefont {M.}~\bibnamefont
  {Ballardini}}, \bibinfo {author} {\bibfnamefont {F.}~\bibnamefont {Finelli}},
  \bibinfo {author} {\bibfnamefont {C.}~\bibnamefont {Fedeli}}, \ and\ \bibinfo
  {author} {\bibfnamefont {L.}~\bibnamefont {Moscardini}},\ }\href {\doibase
  10.1088/1475-7516/2016/10/041} {\bibfield  {journal} {\bibinfo  {journal}
  {JCAP}\ }\textbf {\bibinfo {volume} {1610}},\ \bibinfo {pages} {041}
  (\bibinfo {year} {2016})},\ \Eprint {http://arxiv.org/abs/1606.03747}
  {arXiv:1606.03747 [astro-ph.CO]} \BibitemShut {NoStop}%
%%CITATION = ARXIV:1606.03747;%%
\bibitem [{\citenamefont {Chen}\ \emph
  {et~al.}(2016{\natexlab{b}})\citenamefont {Chen}, \citenamefont {Meerburg},\
  and\ \citenamefont {MŸnchmeyer}}]{Chen:2016zuu}%
  \BibitemOpen
  \bibfield  {author} {\bibinfo {author} {\bibfnamefont {X.}~\bibnamefont
  {Chen}}, \bibinfo {author} {\bibfnamefont {P.~D.}\ \bibnamefont {Meerburg}},
  \ and\ \bibinfo {author} {\bibfnamefont {M.}~\bibnamefont {MŸnchmeyer}},\
  }\href {\doibase 10.1088/1475-7516/2016/09/023} {\bibfield  {journal}
  {\bibinfo  {journal} {JCAP}\ }\textbf {\bibinfo {volume} {1609}},\ \bibinfo
  {pages} {023} (\bibinfo {year} {2016}{\natexlab{b}})},\ \Eprint
  {http://arxiv.org/abs/1605.09364} {arXiv:1605.09364 [astro-ph.CO]}
  \BibitemShut {NoStop}%
%%CITATION = ARXIV:1605.09364;%%
\bibitem [{\citenamefont {Xu}\ \emph {et~al.}(2016)\citenamefont {Xu},
  \citenamefont {Hamann},\ and\ \citenamefont {Chen}}]{Xu:2016kwz}%
  \BibitemOpen
  \bibfield  {author} {\bibinfo {author} {\bibfnamefont {Y.}~\bibnamefont
  {Xu}}, \bibinfo {author} {\bibfnamefont {J.}~\bibnamefont {Hamann}}, \ and\
  \bibinfo {author} {\bibfnamefont {X.}~\bibnamefont {Chen}},\ }\href {\doibase
  10.1103/PhysRevD.94.123518} {\bibfield  {journal} {\bibinfo  {journal} {Phys.
  Rev.}\ }\textbf {\bibinfo {volume} {D94}},\ \bibinfo {pages} {123518}
  (\bibinfo {year} {2016})},\ \Eprint {http://arxiv.org/abs/1607.00817}
  {arXiv:1607.00817 [astro-ph.CO]} \BibitemShut {NoStop}%
%%CITATION = ARXIV:1607.00817;%%
\bibitem [{\citenamefont {Schmidt}\ \emph {et~al.}(2015)\citenamefont
  {Schmidt}, \citenamefont {Chisari},\ and\ \citenamefont
  {Dvorkin}}]{Schmidt:2015xka}%
  \BibitemOpen
  \bibfield  {author} {\bibinfo {author} {\bibfnamefont {F.}~\bibnamefont
  {Schmidt}}, \bibinfo {author} {\bibfnamefont {N.~E.}\ \bibnamefont
  {Chisari}}, \ and\ \bibinfo {author} {\bibfnamefont {C.}~\bibnamefont
  {Dvorkin}},\ }\href {\doibase 10.1088/1475-7516/2015/10/032} {\bibfield
  {journal} {\bibinfo  {journal} {JCAP}\ }\textbf {\bibinfo {volume} {1510}},\
  \bibinfo {pages} {032} (\bibinfo {year} {2015})},\ \Eprint
  {http://arxiv.org/abs/1506.02671} {arXiv:1506.02671 [astro-ph.CO]}
  \BibitemShut {NoStop}%
%%CITATION = ARXIV:1506.02671;%%
\bibitem [{\citenamefont {Mu–oz}\ \emph {et~al.}(2015)\citenamefont {Mu–oz},
  \citenamefont {Ali-Ha•moud},\ and\ \citenamefont
  {Kamionkowski}}]{Munoz:2015eqa}%
  \BibitemOpen
  \bibfield  {author} {\bibinfo {author} {\bibfnamefont {J.~B.}\ \bibnamefont
  {Mu–oz}}, \bibinfo {author} {\bibfnamefont {Y.}~\bibnamefont {Ali-Ha•moud}},
  \ and\ \bibinfo {author} {\bibfnamefont {M.}~\bibnamefont {Kamionkowski}},\
  }\href {\doibase 10.1103/PhysRevD.92.083508} {\bibfield  {journal} {\bibinfo
  {journal} {Phys. Rev.}\ }\textbf {\bibinfo {volume} {D92}},\ \bibinfo {pages}
  {083508} (\bibinfo {year} {2015})},\ \Eprint
  {http://arxiv.org/abs/1506.04152} {arXiv:1506.04152 [astro-ph.CO]}
  \BibitemShut {NoStop}%
%%CITATION = ARXIV:1506.04152;%%
\bibitem [{\citenamefont {Raccanelli}\ \emph {et~al.}(2017)\citenamefont
  {Raccanelli}, \citenamefont {Shiraishi}, \citenamefont {Bartolo},
  \citenamefont {Bertacca}, \citenamefont {Liguori}, \citenamefont {Matarrese},
  \citenamefont {Norris},\ and\ \citenamefont
  {Parkinson}}]{Raccanelli:2015oma}%
  \BibitemOpen
  \bibfield  {author} {\bibinfo {author} {\bibfnamefont {A.}~\bibnamefont
  {Raccanelli}}, \bibinfo {author} {\bibfnamefont {M.}~\bibnamefont
  {Shiraishi}}, \bibinfo {author} {\bibfnamefont {N.}~\bibnamefont {Bartolo}},
  \bibinfo {author} {\bibfnamefont {D.}~\bibnamefont {Bertacca}}, \bibinfo
  {author} {\bibfnamefont {M.}~\bibnamefont {Liguori}}, \bibinfo {author}
  {\bibfnamefont {S.}~\bibnamefont {Matarrese}}, \bibinfo {author}
  {\bibfnamefont {R.~P.}\ \bibnamefont {Norris}}, \ and\ \bibinfo {author}
  {\bibfnamefont {D.}~\bibnamefont {Parkinson}},\ }\href {\doibase
  10.1016/j.dark.2016.10.006} {\bibfield  {journal} {\bibinfo  {journal} {Phys.
  Dark Univ.}\ }\textbf {\bibinfo {volume} {15}},\ \bibinfo {pages} {35}
  (\bibinfo {year} {2017})},\ \Eprint {http://arxiv.org/abs/1507.05903}
  {arXiv:1507.05903 [astro-ph.CO]} \BibitemShut {NoStop}%
%%CITATION = ARXIV:1507.05903;%%
\bibitem [{\citenamefont {Shiraishi}\ \emph
  {et~al.}(2016{\natexlab{a}})\citenamefont {Shiraishi}, \citenamefont
  {Bartolo},\ and\ \citenamefont {Liguori}}]{Shiraishi:2016hjd}%
  \BibitemOpen
  \bibfield  {author} {\bibinfo {author} {\bibfnamefont {M.}~\bibnamefont
  {Shiraishi}}, \bibinfo {author} {\bibfnamefont {N.}~\bibnamefont {Bartolo}},
  \ and\ \bibinfo {author} {\bibfnamefont {M.}~\bibnamefont {Liguori}},\ }\href
  {\doibase 10.1088/1475-7516/2016/10/015} {\bibfield  {journal} {\bibinfo
  {journal} {JCAP}\ }\textbf {\bibinfo {volume} {1610}},\ \bibinfo {pages}
  {015} (\bibinfo {year} {2016}{\natexlab{a}})},\ \Eprint
  {http://arxiv.org/abs/1607.01363} {arXiv:1607.01363 [astro-ph.CO]}
  \BibitemShut {NoStop}%
%%CITATION = ARXIV:1607.01363;%%
\bibitem [{\citenamefont {Moradinezhad~Dizgah}\ and\ \citenamefont
  {Dvorkin}(2017)}]{MoradinezhadDizgah:2017szk}%
  \BibitemOpen
  \bibfield  {author} {\bibinfo {author} {\bibfnamefont {A.}~\bibnamefont
  {Moradinezhad~Dizgah}}\ and\ \bibinfo {author} {\bibfnamefont
  {C.}~\bibnamefont {Dvorkin}},\ }\href@noop {} {\  (\bibinfo {year} {2017})},\
  \Eprint {http://arxiv.org/abs/1708.06473} {arXiv:1708.06473 [astro-ph.CO]}
  \BibitemShut {NoStop}%
%%CITATION = ARXIV:1708.06473;%%
\bibitem [{\citenamefont {Kehagias}\ and\ \citenamefont
  {Riotto}(2017)}]{kehagias:2017cym}%
  \BibitemOpen
  \bibfield  {author} {\bibinfo {author} {\bibfnamefont {A.}~\bibnamefont
  {Kehagias}}\ and\ \bibinfo {author} {\bibfnamefont {A.}~\bibnamefont
  {Riotto}},\ }\href {\doibase 10.1088/1475-7516/2017/07/046} {\bibfield
  {journal} {\bibinfo  {journal} {JCAP}\ }\textbf {\bibinfo {volume} {1707}},\
  \bibinfo {pages} {046} (\bibinfo {year} {2017})},\ \Eprint
  {http://arxiv.org/abs/1705.05834} {arXiv:1705.05834 [hep-th]} \BibitemShut
  {NoStop}%
%%CITATION = ARXIV:1705.05834;%%
\bibitem [{\citenamefont {Vasiliev}(1990)}]{Vas}%
  \BibitemOpen
  \bibfield  {author} {\bibinfo {author} {\bibfnamefont {M.~A.}\ \bibnamefont
  {Vasiliev}},\ }\href {\doibase 10.1016/0370-2693(90)91400-6} {\bibfield
  {journal} {\bibinfo  {journal} {Phys. Lett.}\ }\textbf {\bibinfo {volume}
  {B243}},\ \bibinfo {pages} {378} (\bibinfo {year} {1990})}\BibitemShut
  {NoStop}%
%%CITATION = PHLTA,B243,378;%%
\bibitem [{\citenamefont {Ratra}(1992)}]{Ratra:1991bn}%
  \BibitemOpen
  \bibfield  {author} {\bibinfo {author} {\bibfnamefont {B.}~\bibnamefont
  {Ratra}},\ }\href {\doibase 10.1086/186384} {\bibfield  {journal} {\bibinfo
  {journal} {Astrophys. J.}\ }\textbf {\bibinfo {volume} {391}},\ \bibinfo
  {pages} {L1} (\bibinfo {year} {1992})}\BibitemShut {NoStop}%
%%CITATION = ASJOA,391,L1;%%
\bibitem [{\citenamefont {Martin}\ and\ \citenamefont
  {Yokoyama}(2008)}]{Martin:2007ue}%
  \BibitemOpen
  \bibfield  {author} {\bibinfo {author} {\bibfnamefont {J.}~\bibnamefont
  {Martin}}\ and\ \bibinfo {author} {\bibfnamefont {J.}~\bibnamefont
  {Yokoyama}},\ }\href {\doibase 10.1088/1475-7516/2008/01/025} {\bibfield
  {journal} {\bibinfo  {journal} {JCAP}\ }\textbf {\bibinfo {volume} {0801}},\
  \bibinfo {pages} {025} (\bibinfo {year} {2008})},\ \Eprint
  {http://arxiv.org/abs/0711.4307} {arXiv:0711.4307 [astro-ph]} \BibitemShut
  {NoStop}%
%%CITATION = ARXIV:0711.4307;%%
\bibitem [{\citenamefont {Dulaney}\ and\ \citenamefont
  {Gresham}(2010)}]{Dulaney:2010sq}%
  \BibitemOpen
  \bibfield  {author} {\bibinfo {author} {\bibfnamefont {T.~R.}\ \bibnamefont
  {Dulaney}}\ and\ \bibinfo {author} {\bibfnamefont {M.~I.}\ \bibnamefont
  {Gresham}},\ }\href {\doibase 10.1103/PhysRevD.81.103532} {\bibfield
  {journal} {\bibinfo  {journal} {Phys. Rev.}\ }\textbf {\bibinfo {volume}
  {D81}},\ \bibinfo {pages} {103532} (\bibinfo {year} {2010})},\ \Eprint
  {http://arxiv.org/abs/1001.2301} {arXiv:1001.2301 [astro-ph.CO]} \BibitemShut
  {NoStop}%
%%CITATION = ARXIV:1001.2301;%%
\bibitem [{\citenamefont {Gumrukcuoglu}\ \emph {et~al.}(2010)\citenamefont
  {Gumrukcuoglu}, \citenamefont {Himmetoglu},\ and\ \citenamefont
  {Peloso}}]{Gumrukcuoglu:2010yc}%
  \BibitemOpen
  \bibfield  {author} {\bibinfo {author} {\bibfnamefont {A.~E.}\ \bibnamefont
  {Gumrukcuoglu}}, \bibinfo {author} {\bibfnamefont {B.}~\bibnamefont
  {Himmetoglu}}, \ and\ \bibinfo {author} {\bibfnamefont {M.}~\bibnamefont
  {Peloso}},\ }\href {\doibase 10.1103/PhysRevD.81.063528} {\bibfield
  {journal} {\bibinfo  {journal} {Phys. Rev.}\ }\textbf {\bibinfo {volume}
  {D81}},\ \bibinfo {pages} {063528} (\bibinfo {year} {2010})},\ \Eprint
  {http://arxiv.org/abs/1001.4088} {arXiv:1001.4088 [astro-ph.CO]} \BibitemShut
  {NoStop}%
%%CITATION = ARXIV:1001.4088;%%
\bibitem [{\citenamefont {Watanabe}\ \emph {et~al.}(2010)\citenamefont
  {Watanabe}, \citenamefont {Kanno},\ and\ \citenamefont
  {Soda}}]{Watanabe:2010fh}%
  \BibitemOpen
  \bibfield  {author} {\bibinfo {author} {\bibfnamefont {M.-a.}\ \bibnamefont
  {Watanabe}}, \bibinfo {author} {\bibfnamefont {S.}~\bibnamefont {Kanno}}, \
  and\ \bibinfo {author} {\bibfnamefont {J.}~\bibnamefont {Soda}},\ }\href
  {\doibase 10.1143/PTP.123.1041} {\bibfield  {journal} {\bibinfo  {journal}
  {Prog. Theor. Phys.}\ }\textbf {\bibinfo {volume} {123}},\ \bibinfo {pages}
  {1041} (\bibinfo {year} {2010})},\ \Eprint {http://arxiv.org/abs/1003.0056}
  {arXiv:1003.0056 [astro-ph.CO]} \BibitemShut {NoStop}%
%%CITATION = ARXIV:1003.0056;%%
\bibitem [{\citenamefont {Biagetti}\ \emph {et~al.}(2013)\citenamefont
  {Biagetti}, \citenamefont {Kehagias}, \citenamefont {Morgante}, \citenamefont
  {Perrier},\ and\ \citenamefont {Riotto}}]{Biagetti:2013qqa}%
  \BibitemOpen
  \bibfield  {author} {\bibinfo {author} {\bibfnamefont {M.}~\bibnamefont
  {Biagetti}}, \bibinfo {author} {\bibfnamefont {A.}~\bibnamefont {Kehagias}},
  \bibinfo {author} {\bibfnamefont {E.}~\bibnamefont {Morgante}}, \bibinfo
  {author} {\bibfnamefont {H.}~\bibnamefont {Perrier}}, \ and\ \bibinfo
  {author} {\bibfnamefont {A.}~\bibnamefont {Riotto}},\ }\href {\doibase
  10.1088/1475-7516/2013/07/030} {\bibfield  {journal} {\bibinfo  {journal}
  {JCAP}\ }\textbf {\bibinfo {volume} {1307}},\ \bibinfo {pages} {030}
  (\bibinfo {year} {2013})},\ \Eprint {http://arxiv.org/abs/1304.7785}
  {arXiv:1304.7785 [astro-ph.CO]} \BibitemShut {NoStop}%
%%CITATION = ARXIV:1304.7785;%%
\bibitem [{\citenamefont {Higuchi}(1987)}]{Higuchi:1986py}%
  \BibitemOpen
  \bibfield  {author} {\bibinfo {author} {\bibfnamefont {A.}~\bibnamefont
  {Higuchi}},\ }\href {\doibase 10.1016/0550-3213(87)90691-2} {\bibfield
  {journal} {\bibinfo  {journal} {Nucl. Phys.}\ }\textbf {\bibinfo {volume}
  {B282}},\ \bibinfo {pages} {397} (\bibinfo {year} {1987})}\BibitemShut
  {NoStop}%
%%CITATION = NUPHA,B282,397;%%
\bibitem [{\citenamefont {Ackerman}\ \emph {et~al.}(2007)\citenamefont
  {Ackerman}, \citenamefont {Carroll},\ and\ \citenamefont
  {Wise}}]{Ackerman:2007nb}%
  \BibitemOpen
  \bibfield  {author} {\bibinfo {author} {\bibfnamefont {L.}~\bibnamefont
  {Ackerman}}, \bibinfo {author} {\bibfnamefont {S.~M.}\ \bibnamefont
  {Carroll}}, \ and\ \bibinfo {author} {\bibfnamefont {M.~B.}\ \bibnamefont
  {Wise}},\ }\href {\doibase 10.1103/PhysRevD.75.083502,
  10.1103/PhysRevD.80.069901} {\bibfield  {journal} {\bibinfo  {journal} {Phys.
  Rev.}\ }\textbf {\bibinfo {volume} {D75}},\ \bibinfo {pages} {083502}
  (\bibinfo {year} {2007})},\ \bibinfo {note} {[Erratum: Phys.
  Rev.D80,069901(2009)]},\ \Eprint {http://arxiv.org/abs/astro-ph/0701357}
  {arXiv:astro-ph/0701357 [astro-ph]} \BibitemShut {NoStop}%
%%CITATION = ASTRO-PH/0701357;%%
\bibitem [{\citenamefont {Pullen}\ and\ \citenamefont
  {Kamionkowski}(2007)}]{Pullen:2007tu}%
  \BibitemOpen
  \bibfield  {author} {\bibinfo {author} {\bibfnamefont {A.~R.}\ \bibnamefont
  {Pullen}}\ and\ \bibinfo {author} {\bibfnamefont {M.}~\bibnamefont
  {Kamionkowski}},\ }\href {\doibase 10.1103/PhysRevD.76.103529} {\bibfield
  {journal} {\bibinfo  {journal} {Phys. Rev.}\ }\textbf {\bibinfo {volume}
  {D76}},\ \bibinfo {pages} {103529} (\bibinfo {year} {2007})},\ \Eprint
  {http://arxiv.org/abs/0709.1144} {arXiv:0709.1144 [astro-ph]} \BibitemShut
  {NoStop}%
%%CITATION = ARXIV:0709.1144;%%
\bibitem [{\citenamefont {Pullen}\ and\ \citenamefont
  {Hirata}(2010)}]{Pullen:2010zy}%
  \BibitemOpen
  \bibfield  {author} {\bibinfo {author} {\bibfnamefont {A.~R.}\ \bibnamefont
  {Pullen}}\ and\ \bibinfo {author} {\bibfnamefont {C.~M.}\ \bibnamefont
  {Hirata}},\ }\href {\doibase 10.1088/1475-7516/2010/05/027} {\bibfield
  {journal} {\bibinfo  {journal} {JCAP}\ }\textbf {\bibinfo {volume} {1005}},\
  \bibinfo {pages} {027} (\bibinfo {year} {2010})},\ \Eprint
  {http://arxiv.org/abs/1003.0673} {arXiv:1003.0673 [astro-ph.CO]} \BibitemShut
  {NoStop}%
%%CITATION = ARXIV:1003.0673;%%
\bibitem [{\citenamefont {Ade}\ \emph {et~al.}(2014)\citenamefont {Ade} \emph
  {et~al.}}]{Ade:2013nlj}%
  \BibitemOpen
  \bibfield  {author} {\bibinfo {author} {\bibfnamefont {P.~A.~R.}\
  \bibnamefont {Ade}} \emph {et~al.} (\bibinfo {collaboration} {Planck}),\
  }\href {\doibase 10.1051/0004-6361/201321534} {\bibfield  {journal} {\bibinfo
   {journal} {Astron. Astrophys.}\ }\textbf {\bibinfo {volume} {571}},\
  \bibinfo {pages} {A23} (\bibinfo {year} {2014})},\ \Eprint
  {http://arxiv.org/abs/1303.5083} {arXiv:1303.5083 [astro-ph.CO]} \BibitemShut
  {NoStop}%
%%CITATION = ARXIV:1303.5083;%%
\bibitem [{\citenamefont {Ade}\ \emph {et~al.}(2016{\natexlab{d}})\citenamefont
  {Ade} \emph {et~al.}}]{Ade:2015hxq}%
  \BibitemOpen
  \bibfield  {author} {\bibinfo {author} {\bibfnamefont {P.~A.~R.}\
  \bibnamefont {Ade}} \emph {et~al.} (\bibinfo {collaboration} {Planck}),\
  }\href {\doibase 10.1051/0004-6361/201526681} {\bibfield  {journal} {\bibinfo
   {journal} {Astron. Astrophys.}\ }\textbf {\bibinfo {volume} {594}},\
  \bibinfo {pages} {A16} (\bibinfo {year} {2016}{\natexlab{d}})},\ \Eprint
  {http://arxiv.org/abs/1506.07135} {arXiv:1506.07135 [astro-ph.CO]}
  \BibitemShut {NoStop}%
%%CITATION = ARXIV:1506.07135;%%
\bibitem [{\citenamefont {Shiraishi}\ \emph
  {et~al.}(2016{\natexlab{b}})\citenamefont {Shiraishi}, \citenamefont
  {Muñoz}, \citenamefont {Kamionkowski},\ and\ \citenamefont
  {Raccanelli}}]{Shiraishi:2016omb}%
  \BibitemOpen
  \bibfield  {author} {\bibinfo {author} {\bibfnamefont {M.}~\bibnamefont
  {Shiraishi}}, \bibinfo {author} {\bibfnamefont {J.~B.}\ \bibnamefont
  {Muñoz}}, \bibinfo {author} {\bibfnamefont {M.}~\bibnamefont
  {Kamionkowski}}, \ and\ \bibinfo {author} {\bibfnamefont {A.}~\bibnamefont
  {Raccanelli}},\ }\href {\doibase 10.1103/PhysRevD.93.103506} {\bibfield
  {journal} {\bibinfo  {journal} {Phys. Rev.}\ }\textbf {\bibinfo {volume}
  {D93}},\ \bibinfo {pages} {103506} (\bibinfo {year} {2016}{\natexlab{b}})},\
  \Eprint {http://arxiv.org/abs/1603.01206} {arXiv:1603.01206 [astro-ph.CO]}
  \BibitemShut {NoStop}%
%%CITATION = ARXIV:1603.01206;%%
\bibitem [{\citenamefont {Shiraishi}\ \emph {et~al.}(2017)\citenamefont
  {Shiraishi}, \citenamefont {Sugiyama},\ and\ \citenamefont
  {Okumura}}]{Shiraishi:2016wec}%
  \BibitemOpen
  \bibfield  {author} {\bibinfo {author} {\bibfnamefont {M.}~\bibnamefont
  {Shiraishi}}, \bibinfo {author} {\bibfnamefont {N.~S.}\ \bibnamefont
  {Sugiyama}}, \ and\ \bibinfo {author} {\bibfnamefont {T.}~\bibnamefont
  {Okumura}},\ }\href {\doibase 10.1103/PhysRevD.95.063508} {\bibfield
  {journal} {\bibinfo  {journal} {Phys. Rev.}\ }\textbf {\bibinfo {volume}
  {D95}},\ \bibinfo {pages} {063508} (\bibinfo {year} {2017})},\ \Eprint
  {http://arxiv.org/abs/1612.02645} {arXiv:1612.02645 [astro-ph.CO]}
  \BibitemShut {NoStop}%
%%CITATION = ARXIV:1612.02645;%%
\bibitem [{\citenamefont {Kim}\ and\ \citenamefont
  {Komatsu}(2013)}]{Kim:2013gka}%
  \BibitemOpen
  \bibfield  {author} {\bibinfo {author} {\bibfnamefont {J.}~\bibnamefont
  {Kim}}\ and\ \bibinfo {author} {\bibfnamefont {E.}~\bibnamefont {Komatsu}},\
  }\href {\doibase 10.1103/PhysRevD.88.101301} {\bibfield  {journal} {\bibinfo
  {journal} {Phys. Rev.}\ }\textbf {\bibinfo {volume} {D88}},\ \bibinfo {pages}
  {101301} (\bibinfo {year} {2013})},\ \Eprint {http://arxiv.org/abs/1310.1605}
  {arXiv:1310.1605 [astro-ph.CO]} \BibitemShut {NoStop}%
%%CITATION = ARXIV:1310.1605;%%
\bibitem [{\citenamefont {Bartolo}\ \emph
  {et~al.}(2015{\natexlab{b}})\citenamefont {Bartolo}, \citenamefont
  {Matarrese}, \citenamefont {Peloso},\ and\ \citenamefont
  {Shiraishi}}]{Bartolo:2014hwa}%
  \BibitemOpen
  \bibfield  {author} {\bibinfo {author} {\bibfnamefont {N.}~\bibnamefont
  {Bartolo}}, \bibinfo {author} {\bibfnamefont {S.}~\bibnamefont {Matarrese}},
  \bibinfo {author} {\bibfnamefont {M.}~\bibnamefont {Peloso}}, \ and\ \bibinfo
  {author} {\bibfnamefont {M.}~\bibnamefont {Shiraishi}},\ }\href {\doibase
  10.1088/1475-7516/2015/01/027} {\bibfield  {journal} {\bibinfo  {journal}
  {JCAP}\ }\textbf {\bibinfo {volume} {1501}},\ \bibinfo {pages} {027}
  (\bibinfo {year} {2015}{\natexlab{b}})},\ \Eprint
  {http://arxiv.org/abs/1411.2521} {arXiv:1411.2521 [astro-ph.CO]} \BibitemShut
  {NoStop}%
%%CITATION = ARXIV:1411.2521;%%
\bibitem [{\citenamefont {Naruko}\ \emph {et~al.}(2015)\citenamefont {Naruko},
  \citenamefont {Komatsu},\ and\ \citenamefont {Yamaguchi}}]{Naruko:2014bxa}%
  \BibitemOpen
  \bibfield  {author} {\bibinfo {author} {\bibfnamefont {A.}~\bibnamefont
  {Naruko}}, \bibinfo {author} {\bibfnamefont {E.}~\bibnamefont {Komatsu}}, \
  and\ \bibinfo {author} {\bibfnamefont {M.}~\bibnamefont {Yamaguchi}},\ }\href
  {\doibase 10.1088/1475-7516/2015/04/045} {\bibfield  {journal} {\bibinfo
  {journal} {JCAP}\ }\textbf {\bibinfo {volume} {1504}},\ \bibinfo {pages}
  {045} (\bibinfo {year} {2015})},\ \Eprint {http://arxiv.org/abs/1411.5489}
  {arXiv:1411.5489 [astro-ph.CO]} \BibitemShut {NoStop}%
%%CITATION = ARXIV:1411.5489;%%
\bibitem [{\citenamefont {Ashoorioon}\ \emph {et~al.}(2016)\citenamefont
  {Ashoorioon}, \citenamefont {Casadio},\ and\ \citenamefont
  {Koivisto}}]{Ashoorioon:2016lrg}%
  \BibitemOpen
  \bibfield  {author} {\bibinfo {author} {\bibfnamefont {A.}~\bibnamefont
  {Ashoorioon}}, \bibinfo {author} {\bibfnamefont {R.}~\bibnamefont {Casadio}},
  \ and\ \bibinfo {author} {\bibfnamefont {T.}~\bibnamefont {Koivisto}},\
  }\href {\doibase 10.1088/1475-7516/2016/12/002} {\bibfield  {journal}
  {\bibinfo  {journal} {JCAP}\ }\textbf {\bibinfo {volume} {1612}},\ \bibinfo
  {pages} {002} (\bibinfo {year} {2016})},\ \Eprint
  {http://arxiv.org/abs/1605.04758} {arXiv:1605.04758 [hep-th]} \BibitemShut
  {NoStop}%
%%CITATION = ARXIV:1605.04758;%%
\bibitem [{\citenamefont {Sugiyama}\ \emph {et~al.}(2018)\citenamefont
  {Sugiyama}, \citenamefont {Shiraishi},\ and\ \citenamefont
  {Okumura}}]{Sugiyama:2017ggb}%
  \BibitemOpen
  \bibfield  {author} {\bibinfo {author} {\bibfnamefont {N.~S.}\ \bibnamefont
  {Sugiyama}}, \bibinfo {author} {\bibfnamefont {M.}~\bibnamefont {Shiraishi}},
  \ and\ \bibinfo {author} {\bibfnamefont {T.}~\bibnamefont {Okumura}},\ }\href
  {\doibase 10.1093/mnras/stx2333} {\bibfield  {journal} {\bibinfo  {journal}
  {Mon. Not. Roy. Astron. Soc.}\ }\textbf {\bibinfo {volume} {473}},\ \bibinfo
  {pages} {2737} (\bibinfo {year} {2018})},\ \Eprint
  {http://arxiv.org/abs/1704.02868} {arXiv:1704.02868 [astro-ph.CO]}
  \BibitemShut {NoStop}%
%%CITATION = ARXIV:1704.02868;%%
\bibitem [{\citenamefont {Ramazanov}\ and\ \citenamefont
  {Rubtsov}(2014)}]{Ramazanov:2013wea}%
  \BibitemOpen
  \bibfield  {author} {\bibinfo {author} {\bibfnamefont {S.~R.}\ \bibnamefont
  {Ramazanov}}\ and\ \bibinfo {author} {\bibfnamefont {G.}~\bibnamefont
  {Rubtsov}},\ }\href {\doibase 10.1103/PhysRevD.89.043517} {\bibfield
  {journal} {\bibinfo  {journal} {Phys. Rev.}\ }\textbf {\bibinfo {volume}
  {D89}},\ \bibinfo {pages} {043517} (\bibinfo {year} {2014})},\ \Eprint
  {http://arxiv.org/abs/1311.3272} {arXiv:1311.3272 [astro-ph.CO]} \BibitemShut
  {NoStop}%
%%CITATION = ARXIV:1311.3272;%%
\bibitem [{\citenamefont {Rubtsov}\ and\ \citenamefont
  {Ramazanov}(2015)}]{Rubtsov:2014yua}%
  \BibitemOpen
  \bibfield  {author} {\bibinfo {author} {\bibfnamefont {G.~I.}\ \bibnamefont
  {Rubtsov}}\ and\ \bibinfo {author} {\bibfnamefont {S.~R.}\ \bibnamefont
  {Ramazanov}},\ }\href {\doibase 10.1103/PhysRevD.91.043514} {\bibfield
  {journal} {\bibinfo  {journal} {Phys. Rev.}\ }\textbf {\bibinfo {volume}
  {D91}},\ \bibinfo {pages} {043514} (\bibinfo {year} {2015})},\ \Eprint
  {http://arxiv.org/abs/1406.7722} {arXiv:1406.7722 [astro-ph.CO]} \BibitemShut
  {NoStop}%
%%CITATION = ARXIV:1406.7722;%%
\bibitem [{\citenamefont {Kamionkowski}\ and\ \citenamefont
  {Kovetz}(2014)}]{Kamionkowski:2014wza}%
  \BibitemOpen
  \bibfield  {author} {\bibinfo {author} {\bibfnamefont {M.}~\bibnamefont
  {Kamionkowski}}\ and\ \bibinfo {author} {\bibfnamefont {E.~D.}\ \bibnamefont
  {Kovetz}},\ }\href {\doibase 10.1103/PhysRevLett.113.191303} {\bibfield
  {journal} {\bibinfo  {journal} {Phys. Rev. Lett.}\ }\textbf {\bibinfo
  {volume} {113}},\ \bibinfo {pages} {191303} (\bibinfo {year} {2014})},\
  \Eprint {http://arxiv.org/abs/1408.4125} {arXiv:1408.4125 [astro-ph.CO]}
  \BibitemShut {NoStop}%
%%CITATION = ARXIV:1408.4125;%%
\bibitem [{\citenamefont {Hanson}\ and\ \citenamefont
  {Lewis}(2009)}]{Hanson:2009gu}%
  \BibitemOpen
  \bibfield  {author} {\bibinfo {author} {\bibfnamefont {D.}~\bibnamefont
  {Hanson}}\ and\ \bibinfo {author} {\bibfnamefont {A.}~\bibnamefont {Lewis}},\
  }\href {\doibase 10.1103/PhysRevD.80.063004} {\bibfield  {journal} {\bibinfo
  {journal} {Phys. Rev.}\ }\textbf {\bibinfo {volume} {D80}},\ \bibinfo {pages}
  {063004} (\bibinfo {year} {2009})},\ \Eprint {http://arxiv.org/abs/0908.0963}
  {arXiv:0908.0963 [astro-ph.CO]} \BibitemShut {NoStop}%
%%CITATION = ARXIV:0908.0963;%%
\bibitem [{\citenamefont {Hanson}\ \emph {et~al.}(2010)\citenamefont {Hanson},
  \citenamefont {Lewis},\ and\ \citenamefont {Challinor}}]{Hanson:2010gu}%
  \BibitemOpen
  \bibfield  {author} {\bibinfo {author} {\bibfnamefont {D.}~\bibnamefont
  {Hanson}}, \bibinfo {author} {\bibfnamefont {A.}~\bibnamefont {Lewis}}, \
  and\ \bibinfo {author} {\bibfnamefont {A.}~\bibnamefont {Challinor}},\ }\href
  {\doibase 10.1103/PhysRevD.81.103003} {\bibfield  {journal} {\bibinfo
  {journal} {Phys. Rev.}\ }\textbf {\bibinfo {volume} {D81}},\ \bibinfo {pages}
  {103003} (\bibinfo {year} {2010})},\ \Eprint {http://arxiv.org/abs/1003.0198}
  {arXiv:1003.0198 [astro-ph.CO]} \BibitemShut {NoStop}%
%%CITATION = ARXIV:1003.0198;%%
\bibitem [{\citenamefont {Ma}\ \emph {et~al.}(2011)\citenamefont {Ma},
  \citenamefont {Efstathiou},\ and\ \citenamefont {Challinor}}]{Ma:2011ii}%
  \BibitemOpen
  \bibfield  {author} {\bibinfo {author} {\bibfnamefont {Y.-Z.}\ \bibnamefont
  {Ma}}, \bibinfo {author} {\bibfnamefont {G.}~\bibnamefont {Efstathiou}}, \
  and\ \bibinfo {author} {\bibfnamefont {A.}~\bibnamefont {Challinor}},\ }\href
  {\doibase 10.1103/PhysRevD.89.129901, 10.1103/PhysRevD.83.083005} {\bibfield
  {journal} {\bibinfo  {journal} {Phys. Rev.}\ }\textbf {\bibinfo {volume}
  {D83}},\ \bibinfo {pages} {083005} (\bibinfo {year} {2011})},\ \bibinfo
  {note} {[Erratum: Phys. Rev.D89,no.12,129901(2014)]},\ \Eprint
  {http://arxiv.org/abs/1102.4961} {arXiv:1102.4961 [astro-ph.CO]} \BibitemShut
  {NoStop}%
%%CITATION = ARXIV:1102.4961;%%
\bibitem [{\citenamefont {Tauber}\ \emph {et~al.}(2006)\citenamefont {Tauber},
  \citenamefont {Bersanelli}, \citenamefont {Lamarre}, \citenamefont
  {Efstathiou}, \citenamefont {Lawrence}, \citenamefont {Bouchet},
  \citenamefont {Martinez-Gonzalez}, \citenamefont {Matarrese}, \citenamefont
  {Scott}, \citenamefont {White} \emph {et~al.}}]{Planck:2006aa}%
  \BibitemOpen
  \bibfield  {author} {\bibinfo {author} {\bibfnamefont {J.}~\bibnamefont
  {Tauber}}, \bibinfo {author} {\bibfnamefont {M.}~\bibnamefont {Bersanelli}},
  \bibinfo {author} {\bibfnamefont {J.~M.}\ \bibnamefont {Lamarre}}, \bibinfo
  {author} {\bibfnamefont {G.}~\bibnamefont {Efstathiou}}, \bibinfo {author}
  {\bibfnamefont {C.}~\bibnamefont {Lawrence}}, \bibinfo {author}
  {\bibfnamefont {F.}~\bibnamefont {Bouchet}}, \bibinfo {author} {\bibfnamefont
  {E.}~\bibnamefont {Martinez-Gonzalez}}, \bibinfo {author} {\bibfnamefont
  {S.}~\bibnamefont {Matarrese}}, \bibinfo {author} {\bibfnamefont
  {D.}~\bibnamefont {Scott}}, \bibinfo {author} {\bibfnamefont
  {M.}~\bibnamefont {White}},  \emph {et~al.} (\bibinfo {collaboration}
  {Planck}),\ }\href@noop {} {\  (\bibinfo {year} {2006})},\ \Eprint
  {http://arxiv.org/abs/astro-ph/0604069} {arXiv:astro-ph/0604069 [astro-ph]}
  \BibitemShut {NoStop}%
%%CITATION = ASTRO-PH/0604069;%%
\bibitem [{\citenamefont {Kaiser}(1987)}]{Kaiser:1987qv}%
  \BibitemOpen
  \bibfield  {author} {\bibinfo {author} {\bibfnamefont {N.}~\bibnamefont
  {Kaiser}},\ }\href@noop {} {\bibfield  {journal} {\bibinfo  {journal} {Mon.
  Not. Roy. Astron. Soc.}\ }\textbf {\bibinfo {volume} {227}},\ \bibinfo
  {pages} {1} (\bibinfo {year} {1987})}\BibitemShut {NoStop}%
%%CITATION = MNRAA,227,1;%%
\bibitem [{\citenamefont {Hamilton}(1997)}]{Hamilton:1997zq}%
  \BibitemOpen
  \bibfield  {author} {\bibinfo {author} {\bibfnamefont {A.~J.~S.}\
  \bibnamefont {Hamilton}},\ }in\ \href {\doibase 10.1007/978-94-011-4960-0_17}
  {\emph {\bibinfo {booktitle} {{Ringberg Workshop on Large Scale Structure
  Ringberg, Germany, September 23-28, 1996}}}}\ (\bibinfo {year} {1997})\
  \Eprint {http://arxiv.org/abs/astro-ph/9708102} {arXiv:astro-ph/9708102
  [astro-ph]} \BibitemShut {NoStop}%
%%CITATION = ASTRO-PH/9708102;%%
\bibitem [{\citenamefont {Raccanelli}\ \emph {et~al.}(2016)\citenamefont
  {Raccanelli}, \citenamefont {Bertacca}, \citenamefont {Maartens},
  \citenamefont {Clarkson},\ and\ \citenamefont {Doré}}]{Raccanelli:2013gja}%
  \BibitemOpen
  \bibfield  {author} {\bibinfo {author} {\bibfnamefont {A.}~\bibnamefont
  {Raccanelli}}, \bibinfo {author} {\bibfnamefont {D.}~\bibnamefont
  {Bertacca}}, \bibinfo {author} {\bibfnamefont {R.}~\bibnamefont {Maartens}},
  \bibinfo {author} {\bibfnamefont {C.}~\bibnamefont {Clarkson}}, \ and\
  \bibinfo {author} {\bibfnamefont {O.}~\bibnamefont {Doré}},\ }\href
  {\doibase 10.1007/s10714-016-2076-8} {\bibfield  {journal} {\bibinfo
  {journal} {Gen. Rel. Grav.}\ }\textbf {\bibinfo {volume} {48}},\ \bibinfo
  {pages} {84} (\bibinfo {year} {2016})},\ \Eprint
  {http://arxiv.org/abs/1311.6813} {arXiv:1311.6813 [astro-ph.CO]} \BibitemShut
  {NoStop}%
%%CITATION = ARXIV:1311.6813;%%
\bibitem [{\citenamefont {Tansella}\ \emph {et~al.}(2017)\citenamefont
  {Tansella}, \citenamefont {Bonvin}, \citenamefont {Durrer}, \citenamefont
  {Ghosh},\ and\ \citenamefont {Sellentin}}]{Tansella:2017rpi}%
  \BibitemOpen
  \bibfield  {author} {\bibinfo {author} {\bibfnamefont {V.}~\bibnamefont
  {Tansella}}, \bibinfo {author} {\bibfnamefont {C.}~\bibnamefont {Bonvin}},
  \bibinfo {author} {\bibfnamefont {R.}~\bibnamefont {Durrer}}, \bibinfo
  {author} {\bibfnamefont {B.}~\bibnamefont {Ghosh}}, \ and\ \bibinfo {author}
  {\bibfnamefont {E.}~\bibnamefont {Sellentin}},\ }\href@noop {} {\  (\bibinfo
  {year} {2017})},\ \Eprint {http://arxiv.org/abs/1708.00492} {arXiv:1708.00492
  [astro-ph.CO]} \BibitemShut {NoStop}%
%%CITATION = ARXIV:1708.00492;%%
\bibitem [{\citenamefont {Varshalovich}\ \emph {et~al.}(1988)\citenamefont
  {Varshalovich}, \citenamefont {Moskalev},\ and\ \citenamefont
  {Khersonsky}}]{Varshalovich:1988ye}%
  \BibitemOpen
  \bibfield  {author} {\bibinfo {author} {\bibfnamefont {D.~A.}\ \bibnamefont
  {Varshalovich}}, \bibinfo {author} {\bibfnamefont {A.~N.}\ \bibnamefont
  {Moskalev}}, \ and\ \bibinfo {author} {\bibfnamefont {V.~K.}\ \bibnamefont
  {Khersonsky}},\ }\href@noop {} {\emph {\bibinfo {title} {{Quantum Theory of
  Angular Momentum: Irreducible Tensors, Spherical Harmonics, Vector Coupling
  Coefficients, 3nj Symbols}}}}\ (\bibinfo  {publisher} {World Scientific},\
  \bibinfo {address} {Singapore},\ \bibinfo {year} {1988})\BibitemShut
  {NoStop}%
%%CITATION = INSPIRE-271226;%%
\bibitem [{\citenamefont {Szalay}\ \emph {et~al.}(1998)\citenamefont {Szalay},
  \citenamefont {Matsubara},\ and\ \citenamefont {Landy}}]{Szalay:1997cc}%
  \BibitemOpen
  \bibfield  {author} {\bibinfo {author} {\bibfnamefont {A.~S.}\ \bibnamefont
  {Szalay}}, \bibinfo {author} {\bibfnamefont {T.}~\bibnamefont {Matsubara}}, \
  and\ \bibinfo {author} {\bibfnamefont {S.~D.}\ \bibnamefont {Landy}},\ }\href
  {\doibase 10.1086/311293} {\bibfield  {journal} {\bibinfo  {journal}
  {Astrophys. J.}\ }\textbf {\bibinfo {volume} {498}},\ \bibinfo {pages} {L1}
  (\bibinfo {year} {1998})},\ \Eprint {http://arxiv.org/abs/astro-ph/9712007}
  {arXiv:astro-ph/9712007 [astro-ph]} \BibitemShut {NoStop}%
%%CITATION = ASTRO-PH/9712007;%%
\bibitem [{\citenamefont {Hajian}\ and\ \citenamefont
  {Souradeep}(2003)}]{Hajian:2003qq}%
  \BibitemOpen
  \bibfield  {author} {\bibinfo {author} {\bibfnamefont {A.}~\bibnamefont
  {Hajian}}\ and\ \bibinfo {author} {\bibfnamefont {T.}~\bibnamefont
  {Souradeep}},\ }\href {\doibase 10.1086/379757} {\bibfield  {journal}
  {\bibinfo  {journal} {Astrophys. J.}\ }\textbf {\bibinfo {volume} {597}},\
  \bibinfo {pages} {L5} (\bibinfo {year} {2003})},\ \Eprint
  {http://arxiv.org/abs/astro-ph/0308001} {arXiv:astro-ph/0308001 [astro-ph]}
  \BibitemShut {NoStop}%
%%CITATION = ASTRO-PH/0308001;%%
\bibitem [{\citenamefont {Bolton}\ \emph {et~al.}(2012)\citenamefont {Bolton}
  \emph {et~al.}}]{Bolton:2012hz}%
  \BibitemOpen
  \bibfield  {author} {\bibinfo {author} {\bibfnamefont {A.~S.}\ \bibnamefont
  {Bolton}} \emph {et~al.} (\bibinfo {collaboration} {Cutler Group, LP}),\
  }\href {\doibase 10.1088/0004-6256/144/5/144} {\bibfield  {journal} {\bibinfo
   {journal} {Astron. J.}\ }\textbf {\bibinfo {volume} {144}},\ \bibinfo
  {pages} {144} (\bibinfo {year} {2012})},\ \Eprint
  {http://arxiv.org/abs/1207.7326} {arXiv:1207.7326 [astro-ph.CO]} \BibitemShut
  {NoStop}%
%%CITATION = ARXIV:1207.7326;%%
\bibitem [{\citenamefont {Dawson}\ \emph {et~al.}(2013)\citenamefont {Dawson}
  \emph {et~al.}}]{Dawson:2012va}%
  \BibitemOpen
  \bibfield  {author} {\bibinfo {author} {\bibfnamefont {K.~S.}\ \bibnamefont
  {Dawson}} \emph {et~al.} (\bibinfo {collaboration} {BOSS}),\ }\href {\doibase
  10.1088/0004-6256/145/1/10} {\bibfield  {journal} {\bibinfo  {journal}
  {Astron. J.}\ }\textbf {\bibinfo {volume} {145}},\ \bibinfo {pages} {10}
  (\bibinfo {year} {2013})},\ \Eprint {http://arxiv.org/abs/1208.0022}
  {arXiv:1208.0022 [astro-ph.CO]} \BibitemShut {NoStop}%
%%CITATION = ARXIV:1208.0022;%%
\bibitem [{\citenamefont {Eisenstein}\ \emph {et~al.}(2011)\citenamefont
  {Eisenstein} \emph {et~al.}}]{Eisenstein:2011sa}%
  \BibitemOpen
  \bibfield  {author} {\bibinfo {author} {\bibfnamefont {D.~J.}\ \bibnamefont
  {Eisenstein}} \emph {et~al.} (\bibinfo {collaboration} {SDSS}),\ }\href
  {\doibase 10.1088/0004-6256/142/3/72} {\bibfield  {journal} {\bibinfo
  {journal} {Astron. J.}\ }\textbf {\bibinfo {volume} {142}},\ \bibinfo {pages}
  {72} (\bibinfo {year} {2011})},\ \Eprint {http://arxiv.org/abs/1101.1529}
  {arXiv:1101.1529 [astro-ph.IM]} \BibitemShut {NoStop}%
%%CITATION = ARXIV:1101.1529;%%
\bibitem [{\citenamefont {Ellis}\ \emph {et~al.}(2014)\citenamefont {Ellis}
  \emph {et~al.}}]{Ellis:2012rn}%
  \BibitemOpen
  \bibfield  {author} {\bibinfo {author} {\bibfnamefont {R.}~\bibnamefont
  {Ellis}} \emph {et~al.} (\bibinfo {collaboration} {PFS Team}),\ }\href
  {\doibase 10.1093/pasj/pst019} {\bibfield  {journal} {\bibinfo  {journal}
  {Publ. Astron. Soc. Jap.}\ }\textbf {\bibinfo {volume} {66}},\ \bibinfo
  {pages} {R1} (\bibinfo {year} {2014})},\ \Eprint
  {http://arxiv.org/abs/1206.0737} {arXiv:1206.0737 [astro-ph.CO]} \BibitemShut
  {NoStop}%
%%CITATION = ARXIV:1206.0737;%%
\bibitem [{\citenamefont {Laureijs}\ \emph {et~al.}(2011)\citenamefont
  {Laureijs} \emph {et~al.}}]{Laureijs:2011gra}%
  \BibitemOpen
  \bibfield  {author} {\bibinfo {author} {\bibfnamefont {R.}~\bibnamefont
  {Laureijs}} \emph {et~al.} (\bibinfo {collaboration} {EUCLID}),\ }\href@noop
  {} {\  (\bibinfo {year} {2011})},\ \Eprint {http://arxiv.org/abs/1110.3193}
  {arXiv:1110.3193 [astro-ph.CO]} \BibitemShut {NoStop}%
%%CITATION = ARXIV:1110.3193;%%
\end{thebibliography}%
%\nocite{*}
%%%%%%%%%%%%%%%%%%%%%%%%%%%%%%%%%%%%%%%%%%

\end{document}